\documentclass[apj]{emulateapj}
\usepackage{apjfonts,graphics}
\usepackage{graphicx,floatrow}
\usepackage{subfigure}
\usepackage{natbib}
\usepackage{xcolor}

\newcommand{\scripts}{\fontsize{6.5pt}{0pt}\selectfont}

\def\nar{NewAR}

\begin{document}

\title{Spectral Properties Of Galaxies In Void Regions}
\author{
Chen-Xu~Liu\altaffilmark{1},
Danny~C.~Pan\altaffilmark{1},
Lei~Hao\altaffilmark{1},
Fiona~Hoyle\altaffilmark{2},
Anca~Constantin\altaffilmark{3},
Michael~S.~Vogeley\altaffilmark{4},
}
\email{haol@shao.ac.cn}

\altaffiltext{1}{Key Laboratory for Research in Galaxies and Cosmology of Chinese Acadamy of Sciences, Shanghai Astronomical Observatory, 80 Nandan Road, Shanghai 200030, China}
\altaffiltext{2}{Pontifica Universidad Catolica de Ecuador, 12 de Octubre 1076 y Roca, Quito, Ecuador}
\altaffiltext{3}{James Madison University, Harrisonburg, UA, USA}
\altaffiltext{4}{Department of Physics, Drexel University, Philadelphia, PA 19104, USA}
\begin{abstract}

We present a study of spectral properties of galaxies in underdense large-scale structures, voids. Our void galaxy sample (75,939 galaxies) is selected from the Sloan Digital Sky Survey (SDSS) Data Release 7 (DR7) with $\rm z < 0.107$.  We find that there are no significant differences in the luminosities, stellar masses, stellar populations, and specific star formation rates between void galaxies of specific spectral types and their wall counterparts. However, the fraction of star-forming galaxies in voids is significantly higher ($\ge 9\%$) than that in walls.  Void galaxies, when considering all spectral types, are slightly fainter, less massive, have younger stellar populations and of higher specific star formation rates than wall galaxies. These minor differences are totally caused by the higher fraction of star-forming galaxies in voids. We confirm that AGNs exist in voids,  already found by \cite{co08}, with similar abundance as in walls. Type I AGNs contribute $\sim$ 1\%-2\% of void galaxies, similar to their fraction in walls. The intrinsic [O III] luminosities , spanning over $10^6\ L_{\sun} \sim 10^9\ L_{\sun}$, and Eddington ratios are similar comparing our void AGNs versus wall AGNs. Small scale statistics show that all spectral types of void galaxies are less clustered than their counterparts in walls. Major merger may not be the dominant trigger of black hole accretion in the luminosity range we probe. Our study implies that the growth of black holes relies weakly on large scale structures.

\end{abstract}


\section{Introduction}
\label{sec_intro}

The distribution of galaxies has its large-scale structures. The universe appears as a collection of giant bubble-like voids separated by filaments of galaxies, with the superclusters appearing as occasional relatively dense nodes. Voids are the vast empty spaces with a median effective radius of 17 $h^{-1}$ Mpc, which contain very few, or no, galaxies \citep{pan12}. They were first discovered in 1978 during a pioneering study at the Kitt Peak National Observatory \citep{kitt78}. The low density makes the galaxies less likely to ``see'' or ``greet'' with each other. With few interactions between galaxies, the underdense universe provides a unique laboratory for studying galaxy formation.

Redshift surveys of galaxies enable us to map the universe in three dimensional space. Spectral and photometric analyses of void galaxies in the SDSS DR2 and DR4 samples indicate that void galaxies are bluer, of later type morphologies, and have higher specific star formation rates than galaxies in denser environments \citep{rojas04,rojas05,park07,hoyle12,co08,ricc14}. Despite some clear trends, controversy persists in the literature as to whether or not galaxies in voids differ in their internal properties from similar objects in denser regions. \cite{LF} studied the luminosity function for $\sim1000$  void galaxies generated by \cite{rojas04} from SDSS DR2. They found galaxies located in less dense regions shift to fainter exponential cut-off ($\rm M_r^* =  1.1$ mag) but no evidence for a change in the faint-end slope between voids and walls.  
Different form \cite{LF}, \cite{park07} found that the faint-end (measured only down to $M_r = -18.5$) slope varies significantly with density. \cite{pan12} selected a much larger void galaxy catalog ($\sim10^5$) from SDSS DR7 enabling us to study the spectral properties with higher statistical level and to have a better idea of the possible controversy among previous results of void galaxies.

There are numerous studies address the relation between AGNs and the overdense large scale structures -- the galaxy clusters, as summarized in \cite{Alx12}. \cite{martini06,martini09} and \cite{east07} found that in the low redshift universe (z $<$ 1.3), AGNs are less common in the clusters than in the fields. Contrary to the nearby universe, \cite{lehmer09} studied a protocluster at the redshift about 3.1 with the result that the AGN fraction in the protocluster is 6 times higher than that in fields.  At high redshift, the cold gas supply is more sufficient for AGNs in clusters than in fields. As the clusters virialise, cold gas are heated and can not be accreted efficiently to feed central black hole. Thus galactic activity in the clusters gradually shut down. However, these studies only compared overdense environments with fields. Studies of the occurrence rate of AGNs in less dense regions would therefore be an important complement to better understand the triggering mechanism of AGNs.

\cite{co08} studied the $\sim1000$ magnitude-limited void galaxies selected by \cite{rojas04} from SDSS DR2. They found that AGNs are common in voids, but their occurrence rate and spectral properties are different from those in walls. AGNs are more common in voids, which is mainly caused by the higher fraction of LINER IIs in voids, but only among moderately luminous and massive galaxies. Seyfert IIs are equally abundant in void regions and in wall regions. Void AGNs hosted by moderately luminous and massive galaxies are accreting at equal or lower rates than their wall counterparts, show lower levels of obscuration than in walls, and have similarly aged stellar populations. However, their sample size is too small to give out statistically significant conclusions. There are only 13 Seyferts and 20 LINERs in their $\sim1000$ void galaxy sample. When these AGNs are bined according to their r-band absolute magnitudes for their study, the numbers of galaxies that fall into each bin are even smaller.

In this paper, by using the by far largest void galaxy catalog \citep{pan12}, our goal is to explore how galaxies and, in particular, AGNs are affected by large and small scale environment of galaxies. Section \ref{sec_data} describes the data we use to carry out our study, including the selection method of our void galaxy sample and the classification of galaxies. In Section \ref{sec_results}, we present the spectral properties of galaxies and compare how void galaxies differ from their field counterparts. Section \ref{sec_discuss} discusses the environmental effects on galaxy formation. We summarize our major results in Section \ref{sec_conclusions}.

\section{The Data}
\label{sec_data}

\subsection{The Void Galaxy Sample And The Wall Galaxy Sample}
\label{subsec_sample}

Our analysis are based on the void galaxy catalog generated from the SDSS DR7 Main Galaxy Sample by \cite{pan12}. The SDSS \citep{york2000} is an  imaging and spectroscopic survey using a dedicated 2.5 m telescope at Apache Point, New Mexico. The survey maps one quarter of the entire northern sky. Data Release 7 (DR7)  \citep{aba09}, the seventh major data release of SDSS, covers 8200 square degrees and includes 1,640,960 spectra in total. Among them, 929,555 are galaxies and 121,373 are quasars. These targets are selected from the imaging data via various target selection criteria \citep{strauss2002,rich2002}. In particular, the Main Galaxy Sample has a limit of the r-band Petrosian magnitude at $r=17.77$.

\cite{pan12} used the VoidFinder developed by \cite{hoyle02}, which is a galaxy-based void finder \citep{colberg08}. They start their void-finding process by identifying void galaxy candidates. All volume-limited galaxies ($M_r \le -20.09$, $z<0.107$) with a third nearest neighbor distance $\rm d_3 > 6.3\ h^{-1} Mpc$ are considered to be potential void galaxies and are removed from the total galaxy sample. Only wall galaxies that live in the cosmic filaments and clusters remain. Then these wall galaxies are gridded up in cells of $\rm 5\ h^{-1} Mpc$, allowing them to find voids larger than $\rm 8.5\ h^{-1} Mpc$ in radius. All empty cells are considered to be the centers of potential voids and are grown until bounded by four wall galaxies, reaching maximal spheres. The empty spheres are sorted by descending order of their sizes. The largest empty sphere is the basis of the first void region. If an empty sphere overlaps with an already defined void by $\rm >10\%$, then it is considered to be a subregion of the void, otherwise the sphere becomes the basis of a new void. 

\cite{pan12} did a minimum radius cut at $\rm 10\ h^{-1} Mpc$ for the void regions as they seek to find large-scale structure voids that are dynamically distinct and not small pockets of empty space created by a sparse sample of galaxies. Any field galaxies that lie within a void region are final void galaxies. Deeper surveys may indicate that void galaxies have fine structures and can be classified into isolated galaxies and tendril galaxies \citep{Alpaslan2014}. These structures are at smaller scale, and are still considered as void galaxies in the large scale definition described here. 

With a well defined volume-limited galaxy sample ($M_r \le -20.09$, $z<0.107$), \cite{pan12} detect 1055 voids with radius greater than $\rm 10\ h^{-1} Mpc$. These void regions include 75,939 SDSS DR7 magnitude-limited void galaxies ($z<0.107$, r $<$ 17.77) and 7819 volume-limited void galaxies ($M_r \le -20.09$, $z < 0.107$). The wall galaxy sample is built by removing all void galaxies from the parent SDSS DR7 Main Galaxy Sample at $z < 0.107$ and includes 263,488 magnitude-limited wall galaxies and 107,765 volume-limited wall galaxies. Our wall galaxy sample includes galaxies both in fields and in clusters.

\subsection{Spectral Properties}
\label{subsec_specproper}

\subsubsection{Emission Line Measurements and Spectral Classification}
\label{subsubsec_em}

We adopt methods described in \cite{hao05} to measure the emission lines of both the void galaxies and the wall galaxies.
First, we use the observed equivalent width of the $\rm H_\alpha$ line (in the rest frame) $>$ 3 \AA\ as a criteria to select emission-line galaxies. We consider galaxies not satisfying this criteria as weak or non-emission-line galaxies. We ignore galaxies with weak emission lines for further analysis, since they often need careful stellar subtraction for accurate emission line measurements and spectral classifications. 

For emission-line galaxies, we need to remove the stellar absorption from the spectra so that we can accurately measure the strengths of emission lines and classify the galaxies. This is done by fitting the non-emission-line regions of the spectra with a library of stellar spectra templates, which is constructed by applying the principal component analysis (PCA) method to a sample of pure absorption-line galaxies. The Eigenspectra contain enough information on various absorption features. Thus, they can be used as templates to simulate the stellar components of various galaxies with widely spread metallicities, ages, and velocity dispersions. 
After the stellar subtraction,  main emission lines are measured via Gaussian fits. The $\rm H_\alpha$ and $\rm [N_{\ II}]$ regions are fit with special care because some galaxies may show broad $\rm H_\alpha$ in addition to narrow emission lines.  We fit the $\rm H_\alpha$ and $\rm [N_{\ II}]$ doublet with a four-Gaussian function model first: two for the two $\rm [N_{\ II}]$ lines, one for the broad $\rm H_\alpha$ component and one for the narrow $\rm H_\alpha$ component. The two $\rm H_\alpha$ Gaussian functions have the same central wavelength but different intensities and FWHMs. Then a three-Gaussian model is fit: two for the $\rm [N_{\ II}]$ doublet and one for the narrow $\rm H_\alpha$ emission. The final decision of which model (the three-Gaussian model or the four-Gaussian model) to use for a given galaxy is made by comparing the $\chi^2$ of the two model fits, $\chi_3^2$ and $\chi_4^2$.

\begin{figure*}[!hbt]
\centering
\includegraphics[width=\textwidth]{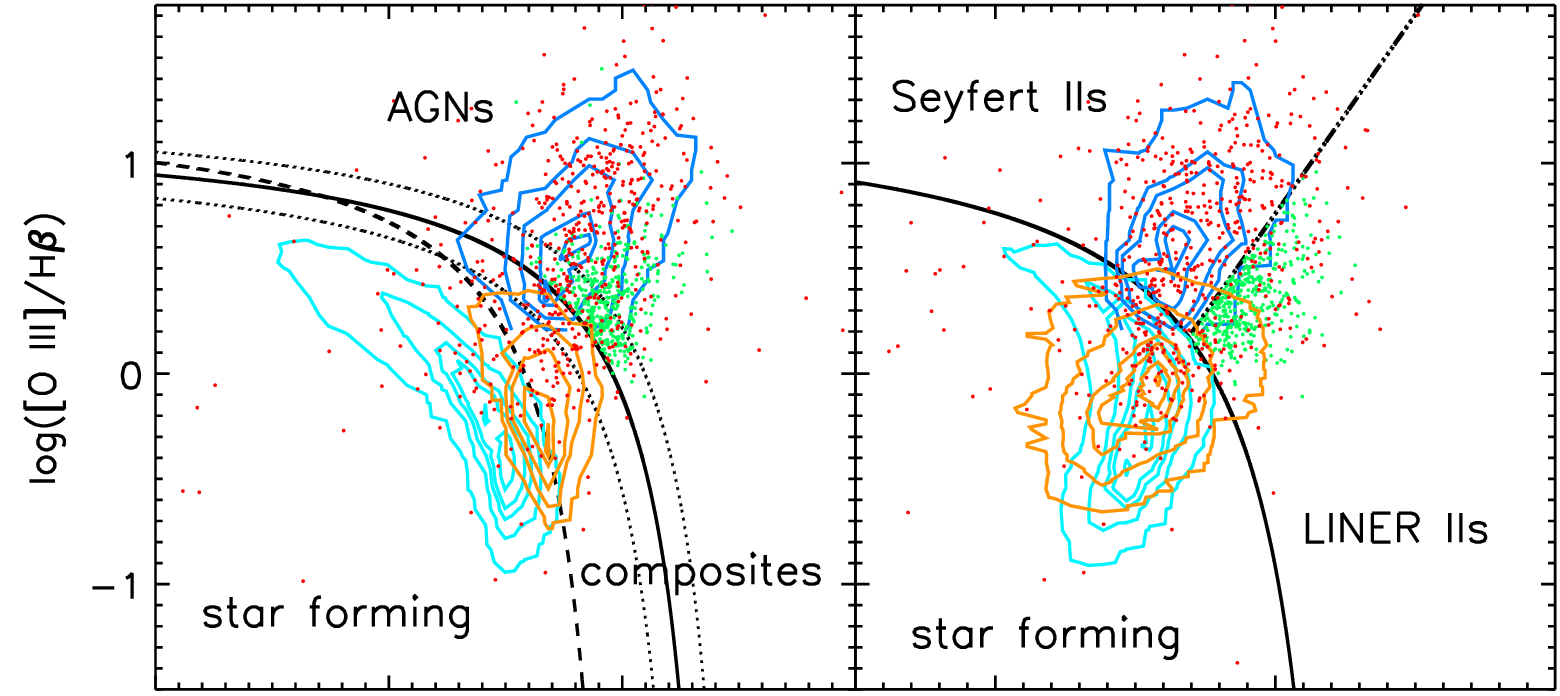}\\
\includegraphics[width=\textwidth]{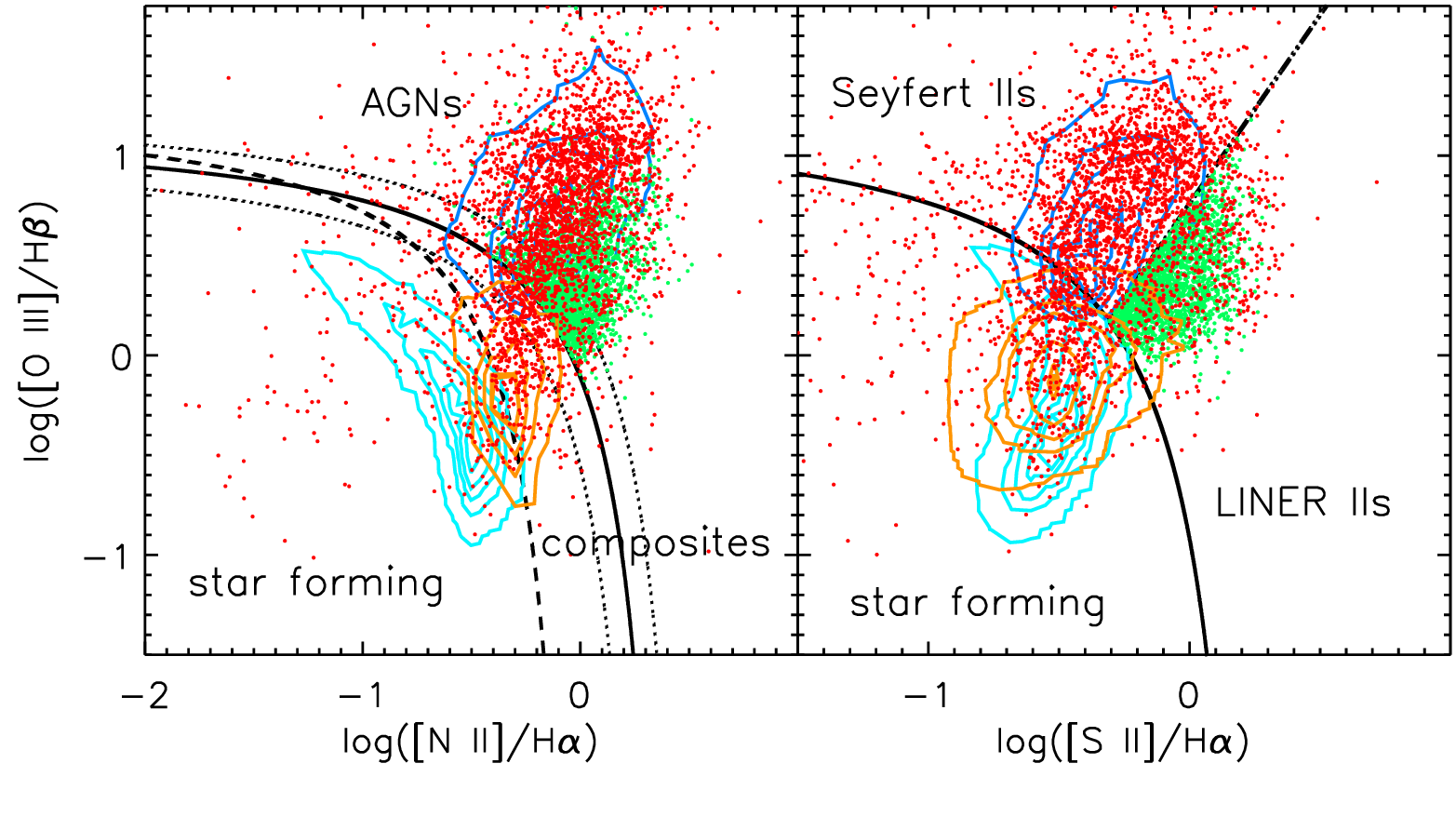}
\caption{Diagnostic diagrams for void galaxies (up) and wall galaxies (down). Dashed and solid hyperbolic curves are based on \cite{ka03b} and \cite{ke01}. Dotted hyperbolic curves are drawn by removing the Ke01 line upward and downward by 0.1 dex, which are used to estimate the errors of AGN fractions. The dot-dot-dashed diagonal segments illustrate the separation between Seyferts and LINERs according to \cite{ke06}. Cyan, blue, and brown isodensity contours show the distributions of star-forming galaxies, Seyfert IIs, and composites, enclosing factors of n of the total number of objects in each class, where n=0.9, 0.6, 0.4, 0.2, and 0.1, starting from the outermost contour. 
LINER IIs are presented by green points. The overlays of red dots on each BPT diagram denote the positions of the line ratios of the narrow emissions of type I AGNs.
\label{bpt}}
\end{figure*}

We first classify type I AGNs from the emission-line galaxies if the $ FWHM(H_\alpha) $ is greater than 1200 km/s.
If two $H_\alpha$ components are needed, the broad $H_\alpha$ is used in this criteria. This criteria is chosen because the distribution of the FWHMs of $H_\alpha$ emissions of all SDSS galaxies is strongly bimodal, with a minimum at 1200 km/s \citep{hao05}. 
For the rest narrow-line galaxies, we classify them as star-forming galaxies, composites, AGNs, and ambiguous galaxies according to their positions on the ``Baldwin, Phillips \& Terlevich'' (BPT) diagrams (\citealt{bpt81,vo87,ke01}, hereafter Ke01; \citealt{ka03b}, hereafter Ka03; \citealt{ke06}, hereafter Ke06; Figure \ref{bpt}). 

The BPT diagrams use flux ratios such as $\rm [O_{\ III}]/H_\beta$, $\rm [N_{\ II}]/H_\alpha$, $\rm [S_{\ II}]/H_\alpha$, and $\rm [O_{\ I}]/H_\alpha$ to identify ionization sources of galaxies. [$\rm O_{\ I}$] emission is usually much weaker than [$\rm N_{\ II}$] emission and [$\rm S_{\ II}$] emission. Thus, in this study we do not use the [$\rm O_{\ I}$] diagram for classifications. The dashed hyperbolic curve in Figure \ref{bpt} is the Ka03 semi-empirical separation between star-forming galaxies and AGNs. The solid hyperbolic curves are the Ke01 boundaries for theoretical star formation models.  
In the [$\rm N_{\ II}$]  diagrams, star-forming galaxies region is under the dashed curve, AGN region is the region above the solid hyperbolic curve, and composites locate in the region between the dashed line and the solid curve. In the [$\rm S_{\ II}$]  diagrams, the region below the solid hyperbolic curve is the star-forming galaxy region, while AGNs are above.

A galaxy that falls in the AGN region in both the [$\rm N_{\ II}$] diagram and the [$\rm S_{\ II}$] diagram is classified as an AGN (blue contours and green dots).  Galaxies that locate in the star-forming region in the two diagrams simultaneously are defined as star-forming galaxies (cyan). If a galaxy shows up in the composite galaxy region in the [$\rm N_{\ II}$] diagram, then it is defined as a composite (brown) no matter where it is in the [$\rm S_{\ II}$] diagram. 

There are some emission-line galaxies that are classified as one type of object in one diagram but another type in the other diagram. These multi-classified galaxies (for example, a galaxy may fall in the star-forming galaxy region in one diagram, but shows up as an AGN in the other diagram) are named as ``ambiguous'' galaxies (Table \ref{classification}).

We further classified AGNs, which are above the red solid curves in both the [$\rm N_{\ II}$] diagram and the [$\rm S_{\ II}$] diagram, into two groups. The dot-dot-dashed diagonal segments in the [$\rm S_{\ II}$]  diagrams in Figure \ref{bpt} are the Ke06 criteria to distinguish Seyferts and LINERs. AGNs that are above the diagonal segment in the [$\rm S_{\ II}$] diagram are identified as Seyfert IIs (blue), while AGNs that are below this diagonal segment are LINER IIs (green). %
Some works argue that the SDSS spectra with LINER like line ratios may better be explained by post-AGB excitation  \citep{erac10,stas08,sarzi10,cid11,cape11,yan12} instead of AGNs. In this paper, all results are not affected by this debate and we include LINERs as AGNs throughout this work for simplicity reasons. 

\subsubsection{Derived Properties}
\label{subsubsec_mpa}

We corrected the emission-line luminosities for extinction using the Balmer decrement and the \cite{cardelli89} reddening curve. We assume an $\rm R_v = A_V/E(B-V) = 3.1$ and an intrinsic $\rm H\alpha/H\beta$ ratio of 2.87 for galaxies dominated by star formation and $\rm H\alpha/H\beta = 3.1$ for AGNs and composites \citep{oster06}. For minor sources ($< 1.8\%$) with the $[H\alpha/H\beta]_{obs}$ less than the intrinsic value due to uncertainties in the emission-line measurement pipeline, we assign them an upper limit of $\rm E(B-V) = 0.001$, which is the minimum E(B-V) of galaxies in our sample with reasonable $[H\alpha/H\beta]_{obs}$.

The absolute r-band magnitude are taken from Korea Institute for Advanced Study Value-Added Galaxy Catalog (KIAS-VAGC) \citep{chk10}. They basically adopted the absolute magnitude in NYU Value-Added Galaxy Catalog (NYU-VAGC) \citep{blanton05}. K-corrections are done following \cite{schlegel98}.

For spectral properties such as stellar masses, the 4000 \AA\ break (D4000) , and star formation rates, we adopt the measurements of the Max-Planck for Astronomy -- John Hopkins University (MPA/JHU) value-added galaxy catalog.
Stellar masses were derived by \cite{salim07} using Monte Carlo stellar population synthesis. They built their stellar library with the code of \cite{bc03}. In their library, they stored a mass-to-light ratio and a photometry in each individual star formation history. Then they weigh each history by fitting the observed photometry with the photometric data in the library. As a result, they can derive the mass-to-light ratio for each galaxy. The model fits provide powerful constraints on the star formation history and metallicity of each galaxy, thus providing a more reliable indicator of mass than assuming a simple M/L.

The 4000 \AA\ break (D4000) is an indicator of the age of the stellar population. The MPA/JHU collaboration measured the 4000 \AA\ breaks according to the definition of \cite{bal99} as the ratio of the average flux density $F_\nu$ in the bands of 4000-4100 and 3850-3950 \AA\ \citep{ka03stm}.

The specific star formation rates (sSFR) are measured according to \cite{brinch04}, with the aperture corrections following \cite{salim07}. In particular the fits to star-forming galaxies are carried out using the \cite{cl01} model. For each galaxy, they fit the observed spectra with single stellar populations (ssp) in the library and used the weight and sSFR of each ssp to estimate the total sSFR.
\ For non-star-forming galaxies, they estimated the sSFRs from the values of D4000 after constructing the likelihood distribution of the sSFR as a function of D4000 using the star-forming galaxy sample. 
 

\section{Results}
\label{sec_results}

\begin{figure} 
\centering
\includegraphics[width=0.98\textwidth]{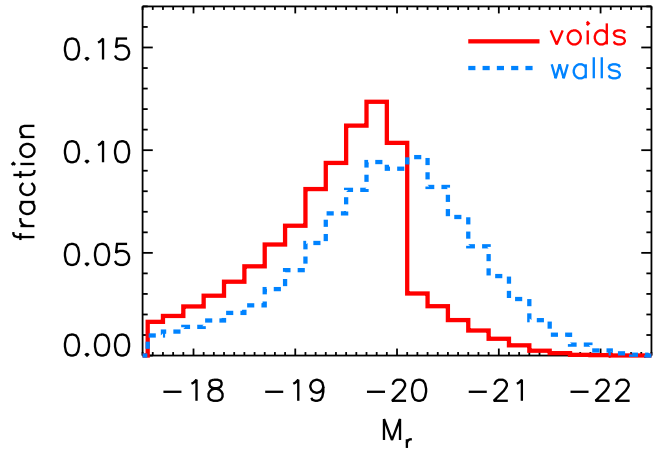}\\
\caption{Distributions of r-band absolute magnitude ($M_r$) for void galaxies (red) and wall galaxies (blue).
\label{Mr_histo_full_mag}}
\end{figure}

Figure \ref{Mr_histo_full_mag} shows the r-band absolute magnitude ($M_r$) distributions for galaxies in voids (red) and in walls (blue). We can see an obvious drop for the number of objects brighter than $M_r \sim -20.1$ in the void sample. This is mainly due to a selection effect.

When generating the void galaxy catalog, \cite{pan12} first constructed the cosmic web by mapping a volume-limited galaxy sample with galaxies of $M_r \le -20.09$, $z < 0.107$. A volume-limited sample is needed in this exercise because it eliminates any bias in the galaxy number density. However, this can create a systematic bias for the magnitude-limited void galaxy sample: brighter galaxies with $M_r \le -20.09$ tend to be identified as wall galaxies when they are on the boundary, since they are the ones that define the cosmic web. Fainter galaxies with $M_r > -20.09$ do not have this tendency. This bias causes the drop at $M_r \sim -20.09$ in the r-band absolute magnitude distribution of the magnitude limited sample (Figure \ref{Mr_histo_full_mag}). Figure \ref{Mr_histo_full} shows that when we separate the magnitude-limited sample into a "bright" ($M_r \le -20.09$) sample and a "faint" ($M_r > -20.09$) sample (Table \ref{3samples}) and only do statistics separately to these two subsamples, we can remove this selection effect from our analysis. Thus, in the following analysis we always consider the two subsamples separately.

\begin{table} 
\caption{Samples \label{3samples}}
\begin{tabular}{c|c|c|c}
\hline\hline sample          &        $M_r$          &  r(Petrosian) &    redshift   \\\hline
magnitude-limited sample       &        --          & $ r < 17.77 $ & $ z \le 0.107 $ \\\hline
   volume-limited sample       & $ M_r\le -20.09 $   &     --        & $ z \le 0.107 $ \\\hline
       faint sample            & $ M_r  > -20.09 $   & $ r < 17.77 $ & $ z \le 0.107 $ \\
\hline\hline
\end{tabular}

\end{table}

\subsection{Object Statistics}
\label{subsec_statistics}

\begin{table} 
\caption{Object Sample Statistics\label{classification}}
\begin{tabular}{c|rr|rr}
\hline\hline(a) magnitude limited sample& & in voids  & & in walls   \\
\hline& N & F(\%) & N & F(\%) \\\hline
   type I                       &      631 &   0.83 &   3153 &   1.20 \\
 Seyfert II                    &   1328 &   1.74 &   5529 &   2.10 \\\hline
AGN without LINER II   &   1959 &   2.58 &   8682 &   3.30  \\
{\scripts (the sum of the above two rows)} &&&& \\ 
  LINER II                     &    433 &   0.57 &   2349 &   0.89 \\\hline
    AGN           &   2392 &   $3.15^{+1.60}_{-0.73}$ &  11031 &   $4.19^{+1.81}_{-0.84}$ \\
{\scripts (the sum of the above two rows)} &&&& \\ 
 composite                     &   6243 &   8.22 &  22548 &   8.56 \\
star-forming                   &  37764 &  49.73 &  81807 &  31.05 \\
 ambiguous                     &   3608 &   4.75 &   5580 &   2.12 \\\hline
  emission                     &  50007 &  65.85 & 120966 &  45.91 \\
{\scripts (the sum of the above four rows)} &&&& \\ 
no emission                    &  25932 &  34.15 & 142522 &  54.09 \\\hline
   total                       &  75939 & 100.00 & 263488 & 100.00 \\
{\scripts (the sum of the above two rows)} &&&& \\ \hline\hline
\end{tabular}

\begin{tabular}{crrrr}\\
\end{tabular}

\begin{tabular}{c|rr|rr}
\hline\hline(b) volume limited sample& &in voids  & & in walls   \\\hline
& N & F(\%) & N & F(\%) \\\hline
   type I                       &    174 &   2.23 &   2157 &   2.00 \\
 Seyfert II                    &    258 &   3.30 &   3040 &   2.82 \\\hline
AGN without LINER II &    432 &   5.53 &   5197 &   4.82 \\
{\scripts (the sum of the above two rows)} &&&& \\ 
  LINER II                     &    101 &   1.29 &   1444 &   1.34 \\\hline
    AGN                        &    533 &   $6.82^{+2.55}_{-1.05}$ &   6641 &   $6.16^{+2.27}_{-0.98}$ \\
{\scripts (the sum of the above two rows)} &&&& \\ 
 composite                     &    971 &  12.42 &  10728 &   9.95 \\
star-forming                   &   1834 &  23.46 &  16575 &  15.38 \\
 ambiguous                     &     51 &   0.65 &    596 &   0.55 \\\hline
  emission                     &   3389 &  43.34 &  34540 &  32.05 \\
{\scripts (the sum of the above four rows)} &&&& \\ 
 no emission                   &   4430 &  56.66 &  73225 &  67.95 \\\hline
   total                       &   7819 & 100.00 & 107765 & 100.00 \\
{\scripts (the sum of the above two rows)} &&&& \\ \hline\hline
\end{tabular}

\begin{tabular}{crrrr}\\
\end{tabular}

\begin{tabular}{c|rr|rr}
\hline\hline(c) faint sample& &in voids  & & in walls    \\\hline
& N & F(\%) & N & F(\%) \\\hline
   type I                        &    457  &   0.67 &     996 &   0.64 \\
 Seyfert II                    &   1070 &   1.57 &   2489 &   1.60 \\\hline
AGN without LINER II &   1527 &   2.24 &   3485 &   2.24 \\
{\scripts (the sum of the above two rows)} &&&& \\  
 LINER II                      &    332  &   0.48 &     905 &   0.58 \\\hline
    AGN                        &   1859 &   $2.73^{+1.49}_{-0.70}$ &   4390 &   $2.82^{+1.50}_{-0.73}$ \\
{\scripts (the sum of the above two rows)} &&&& \\ 
 composite                     &   5272 &   7.74 &  11820 &   7.59 \\
star-forming                   &  35930 &  52.75 &  65232 &  41.89 \\
 ambiguous                     &   3557 &   5.22 &   4984 &   3.20 \\\hline
  emission                     &  46618 &  68.44 &  86426 &  55.50 \\
{\scripts (the sum of the above four rows)} &&&& \\ 
 no emission                   &  21502 &  31.56 &  69297 &  44.50 \\\hline
   total                       &  68120 & 100.00 & 155723 & 100.00 \\
{\scripts (the sum of the above two rows)} &&&& \\ \hline\hline
\end{tabular}

\tablenotetext{0}{\\
$``total" = ``emission" + ``no\ emission"$ \\ 
$ ``emission" = ``AGN" + ``composite" + ``star\-forming" + ``ambiguous"$\\
$ ``AGN" = ``AGN\ without\ LINER\ II" + ``LINER\ II"$\\
$ ``AGN\ without\ LINER\ II" = ``type\ I" + ``Seyfert\ II"$. \\
Errors of AGN fractions are estimated by removing the Ke01 line upward and downward by 0.1 dex.}
\end{table}

We present the object statistics for void galaxies and their wall counterparts in Table \ref{classification}. The percentages of strong emitters (EW($H_\alpha) >$ 3 \AA) and their subclasses of galaxies (such as AGNs, star-forming galaxies, composites, and ambiguous galaxies) are the fractions of objects relative to the whole void and wall galaxy samples. We found that there is a significant partition difference between void galaxies and wall galaxies: emission-line galaxies is almost 30\% more abundant in void regions than in walls for both the volume-limited sample and the faint sample. The higher occurrence rate of emission-line galaxies in less dense regions is mainly contributed by the high fraction of star-forming galaxies in voids. There are $\sim$ 30\% more star-forming galaxies in voids than in walls. In Table \ref{classification}, we also present the object statistics for the magnitude-limited sample as a whole. The trends are similar.

\cite{co08} also found that void galaxies have $\sim$ 30\% more emission-line galaxies, and this is also mainly contributed by the number excess of star-forming galaxies.

\cite{ricc14} studied the faint void galaxy catalog ( $\sim 6000$ void galaxies,  $M_r > -20.17$, $0.01 \le z \le 0.12$) generated by \cite{varela12} from the SDSS DR7 sample. 
Galaxies are found naturally separated into two branches with different specific star formation rates in the star formation rate versus stellar mass diagram. They define star-forming galaxies as the branch with higher specific star formation rates and the other branch as passive galaxies. 
With this definition, they also found that the star forming galaxy fraction in voids (0.92) is higher than that (0.85) in the control sample, even though they used a different definition of star-forming galaxies from our classification scheme.

Objects of other spectral types show about equal frequencies in the underdense regions compared to the wall regions. Active black hole systems (type Is $+$ Seyfert IIs $+$ LINER IIs) are of similar abundances in voids versus in walls. In the volume-limited sample, the AGN fraction is $\sim$ 6\% both in voids and in walls, and in faint sample, the fraction is $\sim$ 3\% in both environments. Errors of AGN fractions listed in Table \ref{classification} are estimated by removing the Ke01 line upward and downward by 0.1 dex. These errors can also serve as an approximation for fractions of composites and star-forming galaxies. The fraction difference of AGNs in the two different environments is smaller than 1\% in all the three samples (Table \ref{3samples}), which is well within the statistical errors shown in Table \ref{classification}. The percentages of individual subclasses of AGNs do not diverge significantly with the void galaxy sample and the wall galaxy sample. 
Whether LINER IIs are included as AGNs or not does not affect our result that the difference of AGN fractions in voids and in walls is minor. \cite{co08} also found that Seyfert IIs were equally represented in voids and walls. 

\subsection{Spectral Properties of Void Galaxies}
\label{subsec_gal}

\subsubsection{Luminosities and Stellar Masses}
\label{subsubsec_Mr_stmass}

\begin{figure} 
\centering
\subfigure[volume-limited sample]{
\includegraphics[width=0.98\textwidth]{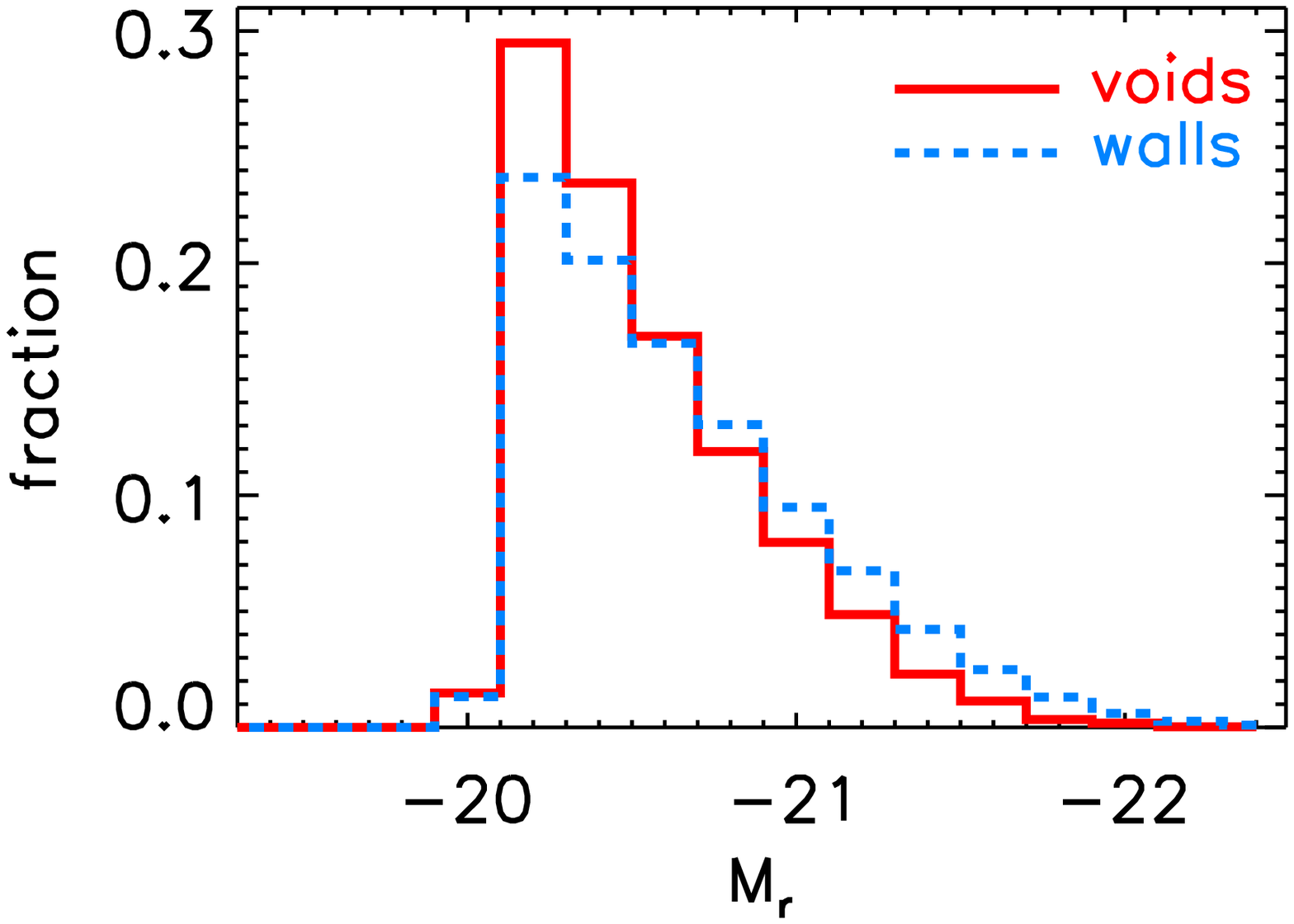}}\\
\subfigure[faint sample]{
\includegraphics[width=0.98\textwidth]{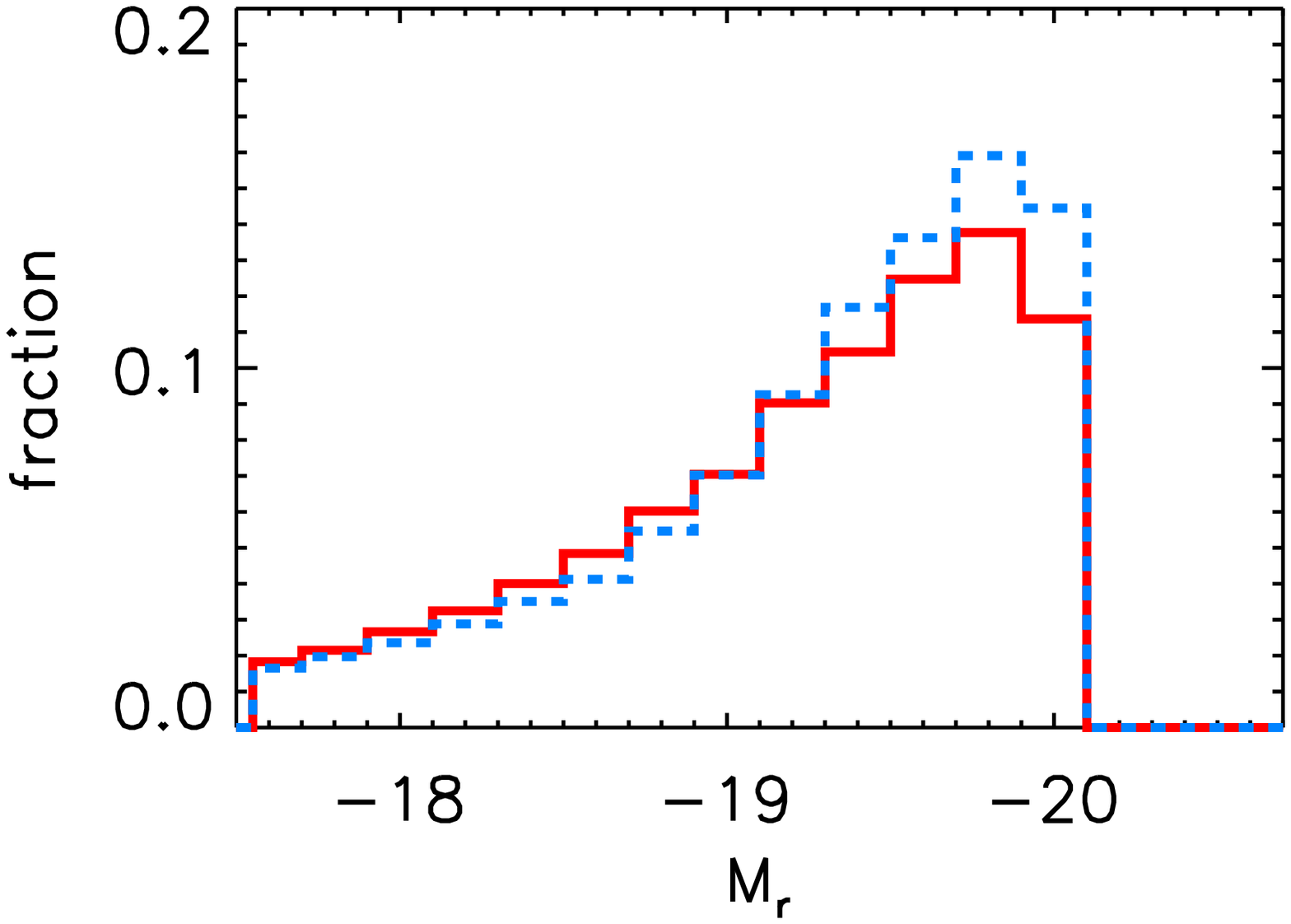}}
\caption{Distributions of r-band absolute magnitude ($M_r$) of void galaxies (red) and wall galaxies (blue) for the volume-limited sample (up) and for the faint sample (down).
\label{Mr_histo_full}}
\end{figure}

\begin{figure} 
\centering
\subfigure[volume-limited sample]{
\includegraphics[width=\textwidth]{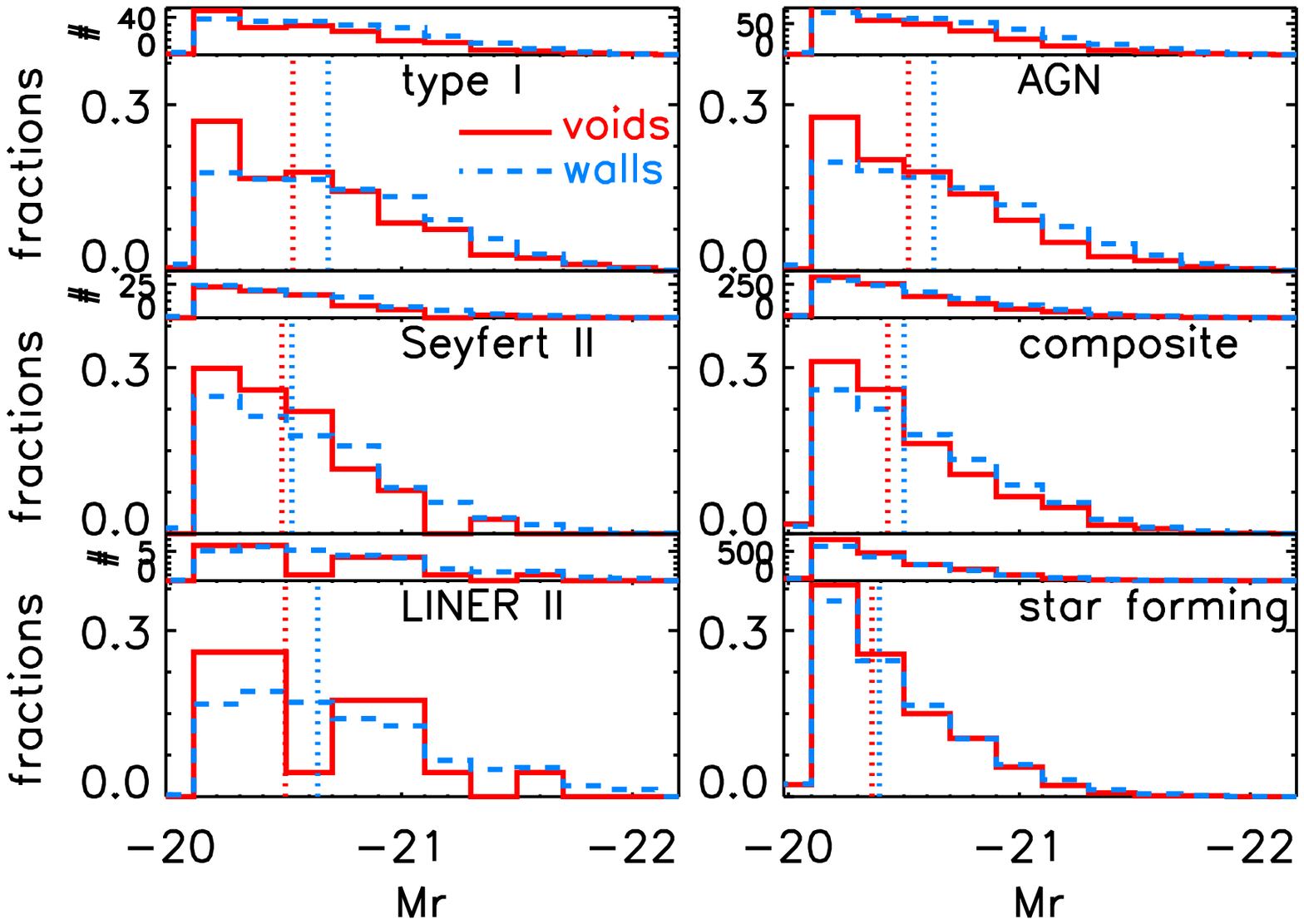}}\\
\subfigure[faint sample]{
\includegraphics[width=\textwidth]{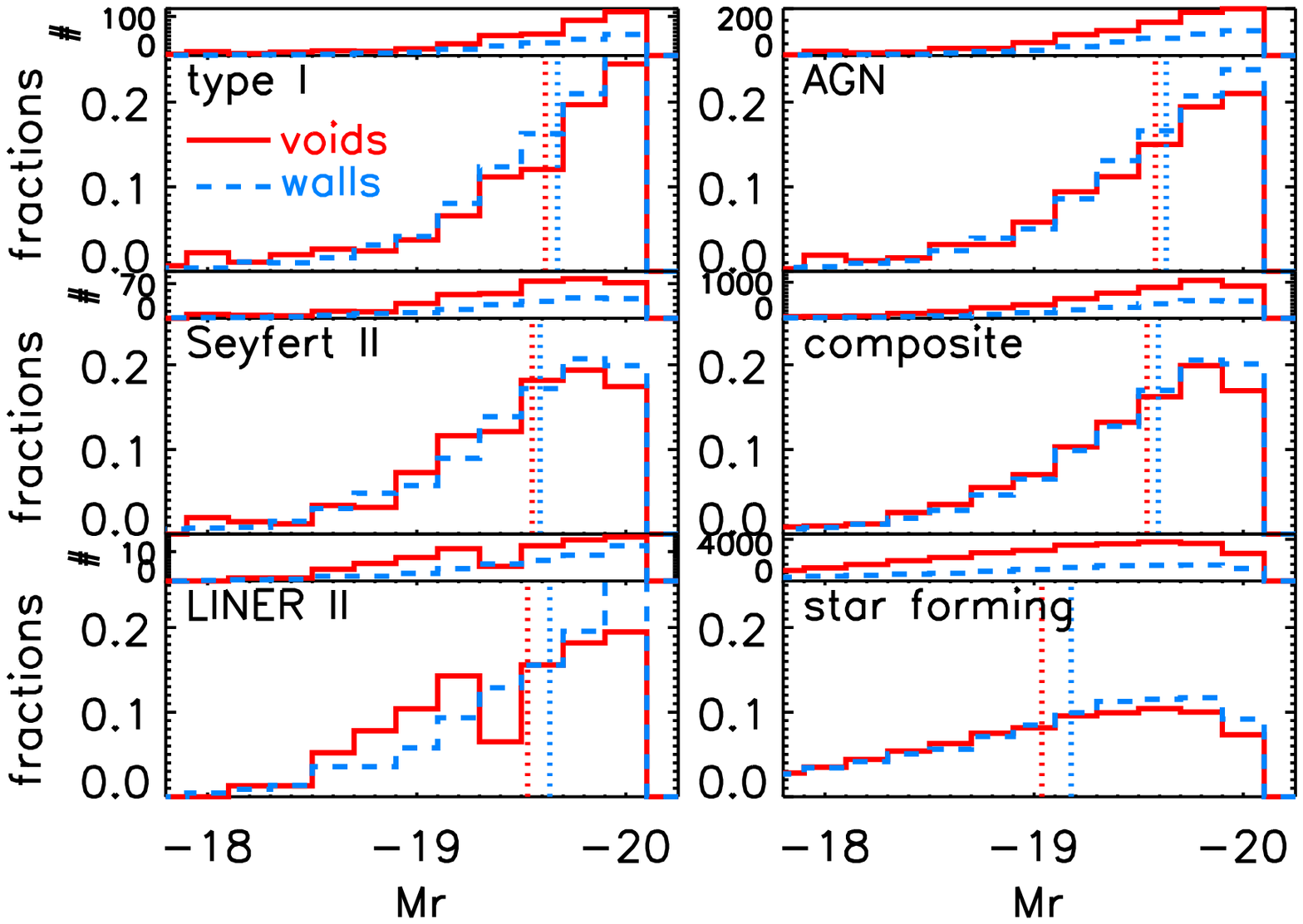}}
\caption{Distributions of r-band absolute magnitude ($M_r$) of each spectral type of galaxies for void galaxies (red) and their wall counterparts (blue). Dotted lines are median values of each sub sample, red for the void galaxies and blue for the wall galaxies. We divide the number counts of galaxies in walls by a factor of 10 for the volume-limited sample and by a factor of 5 for the faint sample for presentation purpose. Note: AGNs = type Is + Seyfert IIs + LINER IIs. 
\label{Mr_distr}}
\end{figure}

Figure \ref{Mr_histo_full} illurstrates the r-band absolute magnitude distributions for the volume-limited sample and the faint sample. For both of the two samples, we see a small systematic faint shift ($<$ 0.2 mag) from the wall galaxies to the void ones. There are some previous works that also found that galaxies in less dense regions are less luminous and less massive than their counterparts that live in denser regions \citep[e.g.,][]{LF,ka04,MF}.

We further study the luminosity distributions for individual spectral type of galaxies in voids and in walls in Figure \ref{Mr_distr}. In both the volume-limited sample and the faint sample,  the luminosity distributions of individual spectral type (AGNs, composites, and star-forming galaxies) of void galaxies and their wall counterparts are indistinguishable. The separations of the luminosity medians are small for all spectral types. This is confirmed by the k-s tests (Table \ref{kstests}).

However, our result that each class of void galaxies are equally luminous compared to their wall counterparts differs from what was shown in \cite{co08}. They also studied the luminosities of detailed spectral types of void galaxies. They found that Seyfert IIs, LINER IIs, composites, and star-forming galaxies in void regions are clearly less luminous than their wall counterparts by $\sim$ 0.5 mag. We think this difference is due to the selection bias we mentioned at the beginning of Section \ref{sec_results}. The void galaxies in \cite{co08} are selected with an earlier version of the voidfinder as \cite{pan12}, which can also suffer the selection bias: For the magnitude-limited sample , more luminous galaxies tend to be identified as wall galaxies as they are the ones that define the cosmic web.  Unfortunately, \cite{co08} did their study only with the magnitude-limited sample due to the small sample size. If we consider only the our magnitude-limited sample in our analysis as did in \cite{co08}, we can also see that void galaxies are less luminous by $\sim$ 0.5 mag (Table \ref{kstests}).

The object statistics studied in Section \ref{subsec_statistics} tell us that the partition between the void galaxies and the wall galaxies are significantly different, with more star-forming galaxies (about 30\% difference) reside in the under dense universe. Figure \ref{Mr_distr} shows that the median luminosities of star-forming galaxies are smaller compared to other types of emission-line galaxies regardless of the large-scale environment, as confirmed by Table \ref{kstests}. This trend has been found for line-emitting galaxies in fields \citep{ho1997}. When considering all spectral types together (Figure \ref{Mr_histo_full}), we do see that void galaxies are slightly less luminous than wall galaxies ($<$ 0.2 mag). This minor luminosity difference is totally caused by the fact the abundance of star-forming galaxies in voids is considerably higher than that in walls.

\begin{figure} 
\centering
\subfigure[volume-limited sample]{
\includegraphics[width=0.98\textwidth]{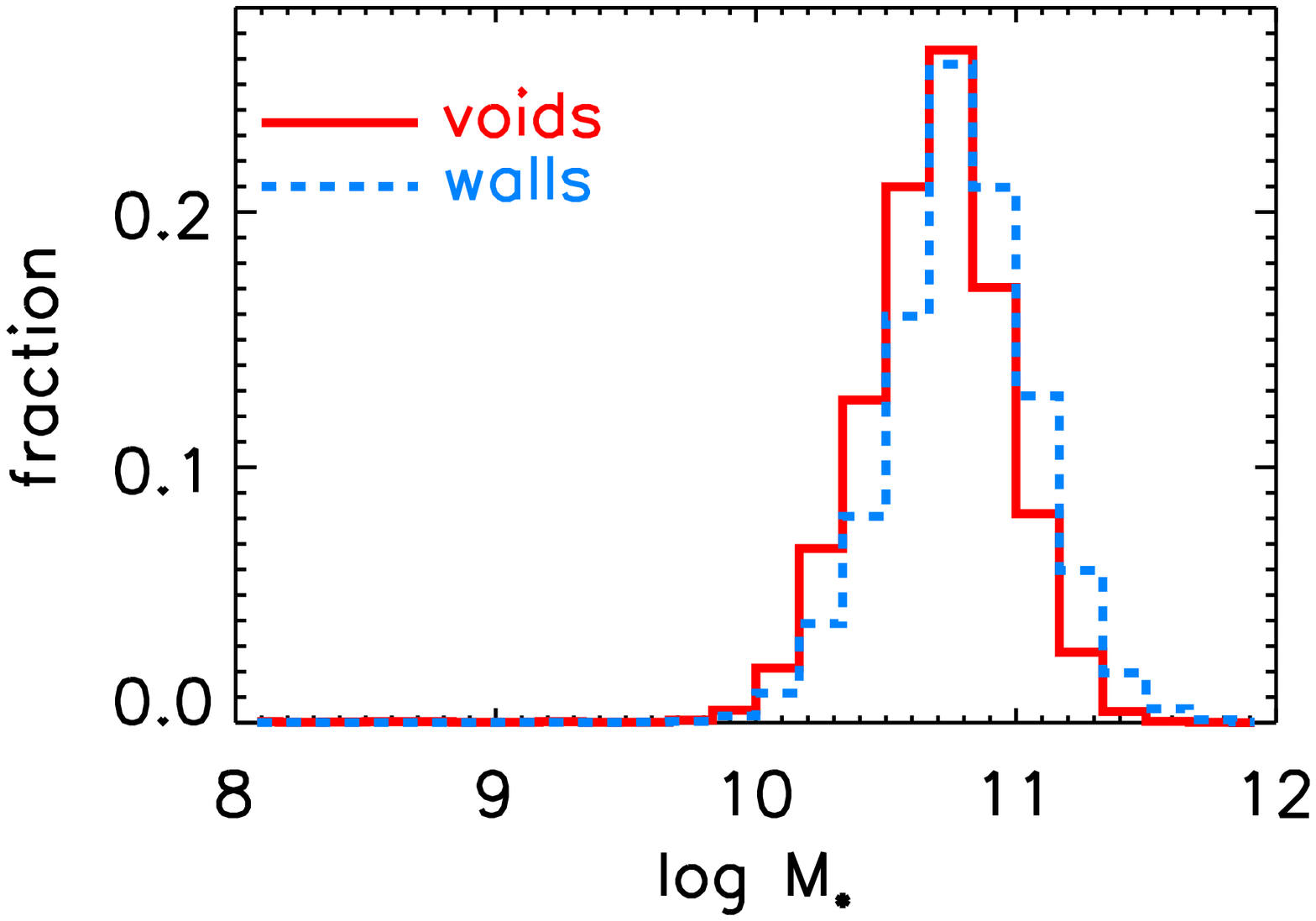}}\\
\subfigure[faint sample]{
\includegraphics[width=0.98\textwidth]{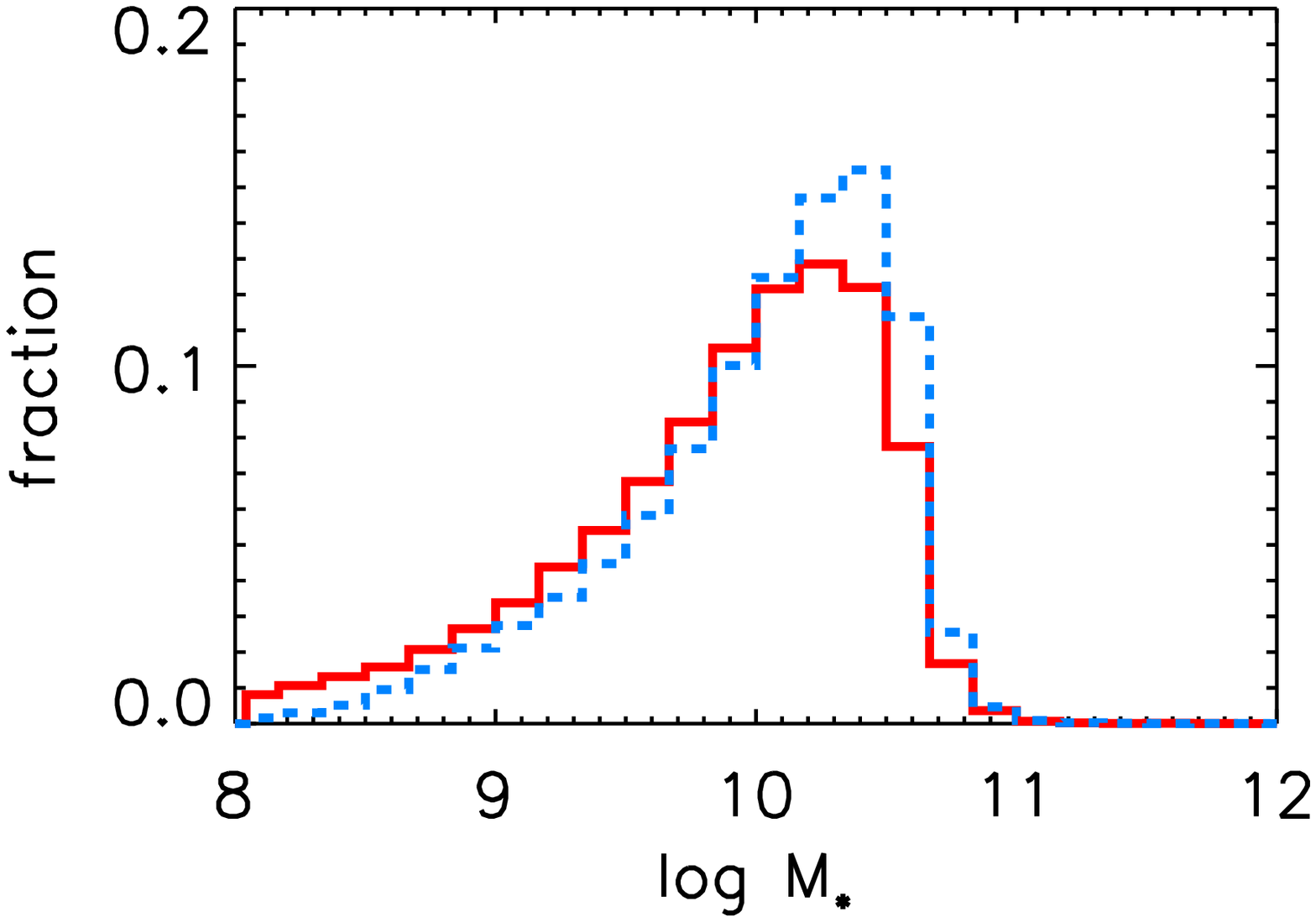}}
\caption{Distributions of stellar masses ($M_*$) of void galaxies (red) and wall galaxies (blue) for the volume-limited sample (up) and for the faint sample (down).
\label{stmass_histo_full}}
\end{figure}

\begin{figure} 
\centering
\subfigure[volume-limited sample]{
\includegraphics[width=\textwidth]{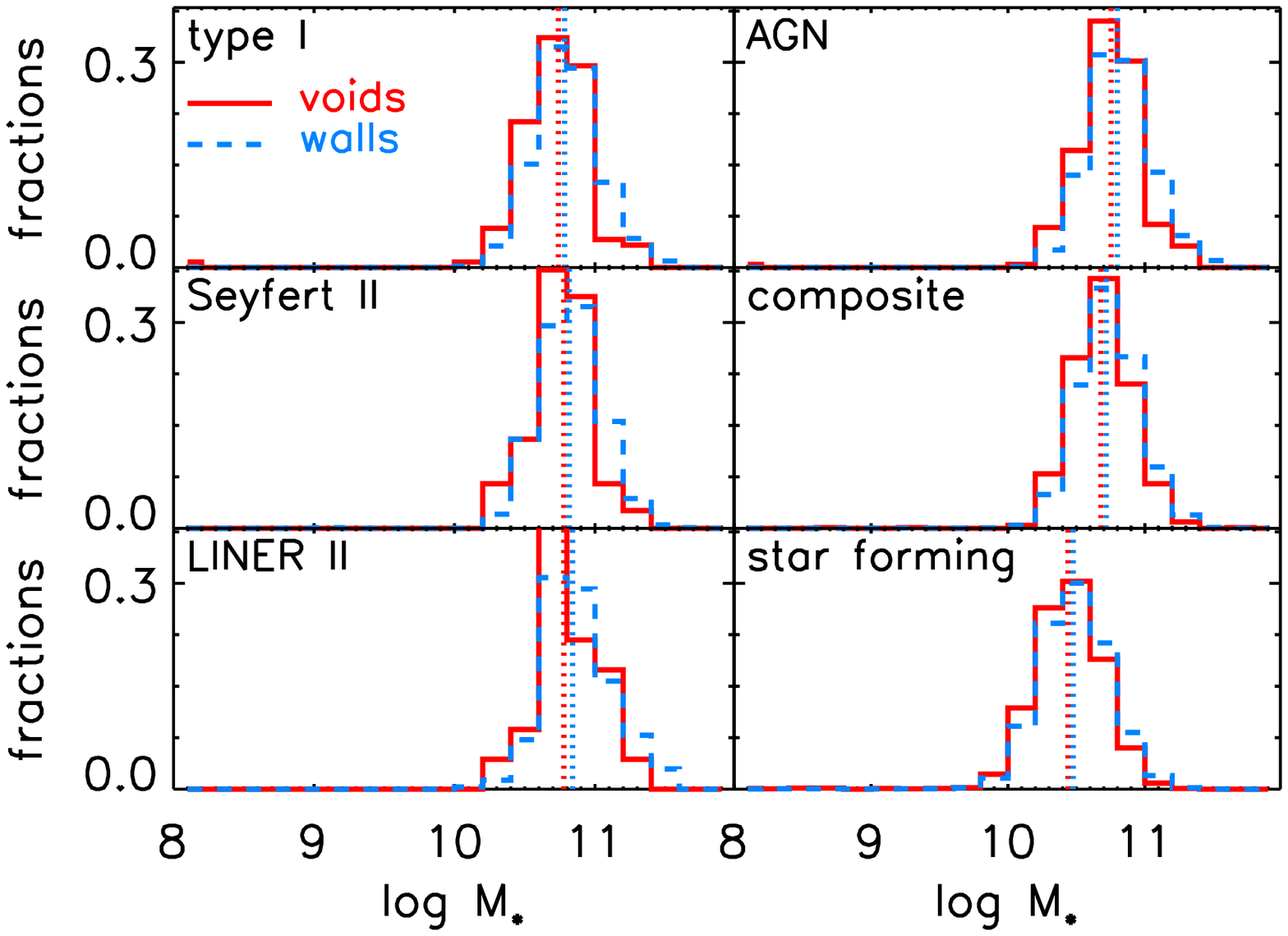}}\\
\subfigure[faint sample]{
\includegraphics[width=\textwidth]{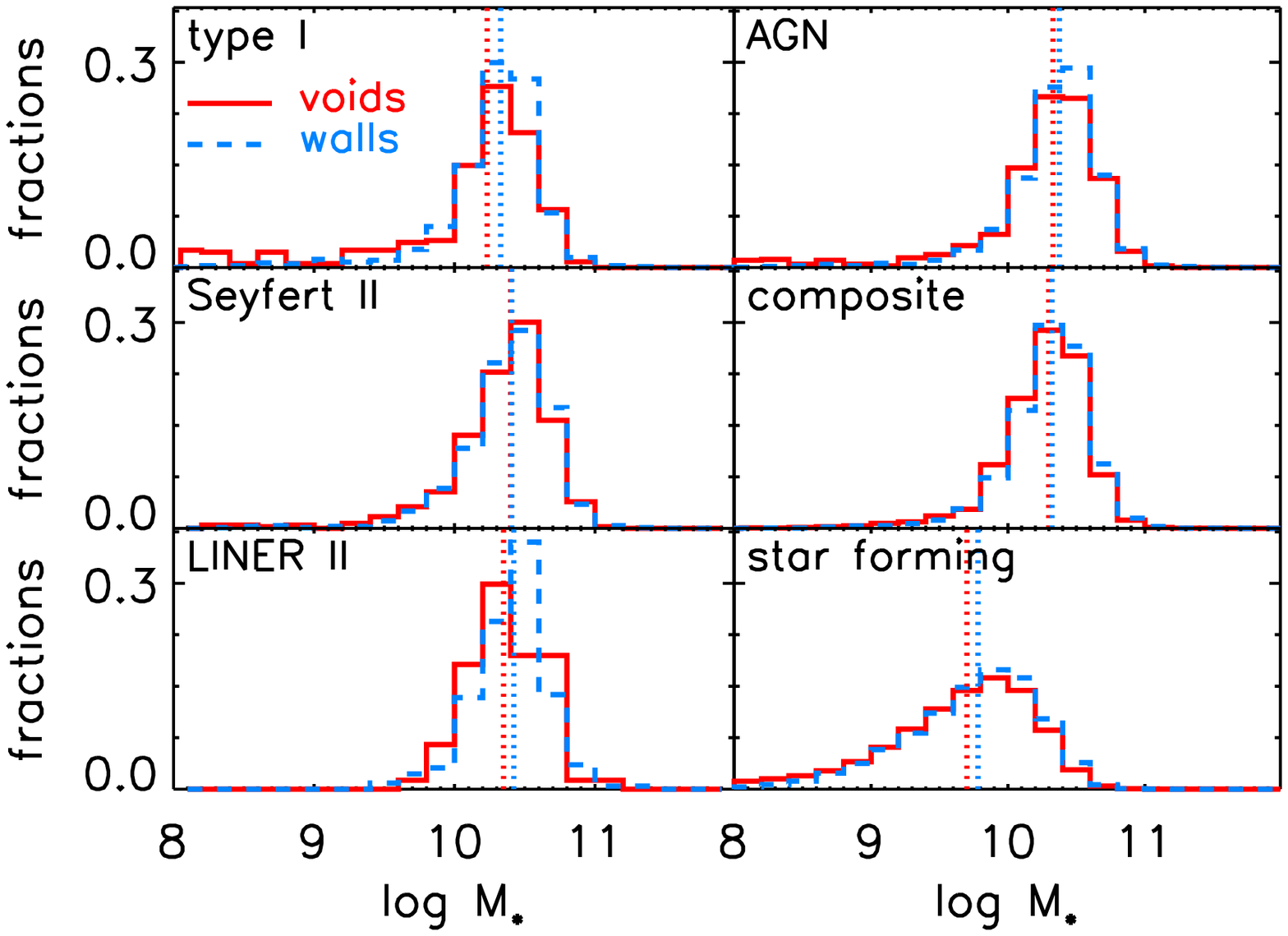}}
\caption{Distributions of stellar masses ($M_*$)  for each spectral type of void galaxies (red) and their wall counterparts (blue) in the volume-limited sample (up) and in the faint sample (down). Dotted lines are median values of each sub sample, red for void galaxies and blue for wall galaxies.
Note: AGNs = type Is + Seyfert IIs + LINER IIs.
\label{stmass_distr}}
\end{figure}

Figure \ref{stmass_histo_full} shows the stellar mass distributions for the volume-limited sample and the faint sample. Again, we can see that void galaxies are slightly less massive than wall galaxies. We also study the stellar mass distributions for individual spectral type of galaxies in voids and in walls in Figure \ref{stmass_distr} as we did for the $M_r$ study. Consistent with our results of the luminosities of individual types of void galaxies, the stellar masses of AGNs, composites, and star-forming galaxies in voids do not significantly diverge from their wall counterparts. Again, the results are confirmed by the k-s test (Table \ref{kstests}).

Consistent with what we found for the $M_r$ distributions, Figure \ref{stmass_distr} and Table \ref{kstests} show us that star-forming galaxies are less massive than other types of galaxies. The over abundance of star-forming galaxies again results in that void galaxies, when considering all spectral types together (Figure \ref{stmass_histo_full}), are slightly less massive than wall galaxies ($<$ 0.1 dex). This minor stellar mass difference is totally caused by the higher occurence rate of star-forming galaxies in less dense regions.

\subsubsection{Stellar populations and Star Formation Rates}
\label{subsubsec_sfr}

\begin{figure} 
\centering
\subfigure[volume-limited sample]{
\includegraphics[width=0.8\textwidth]{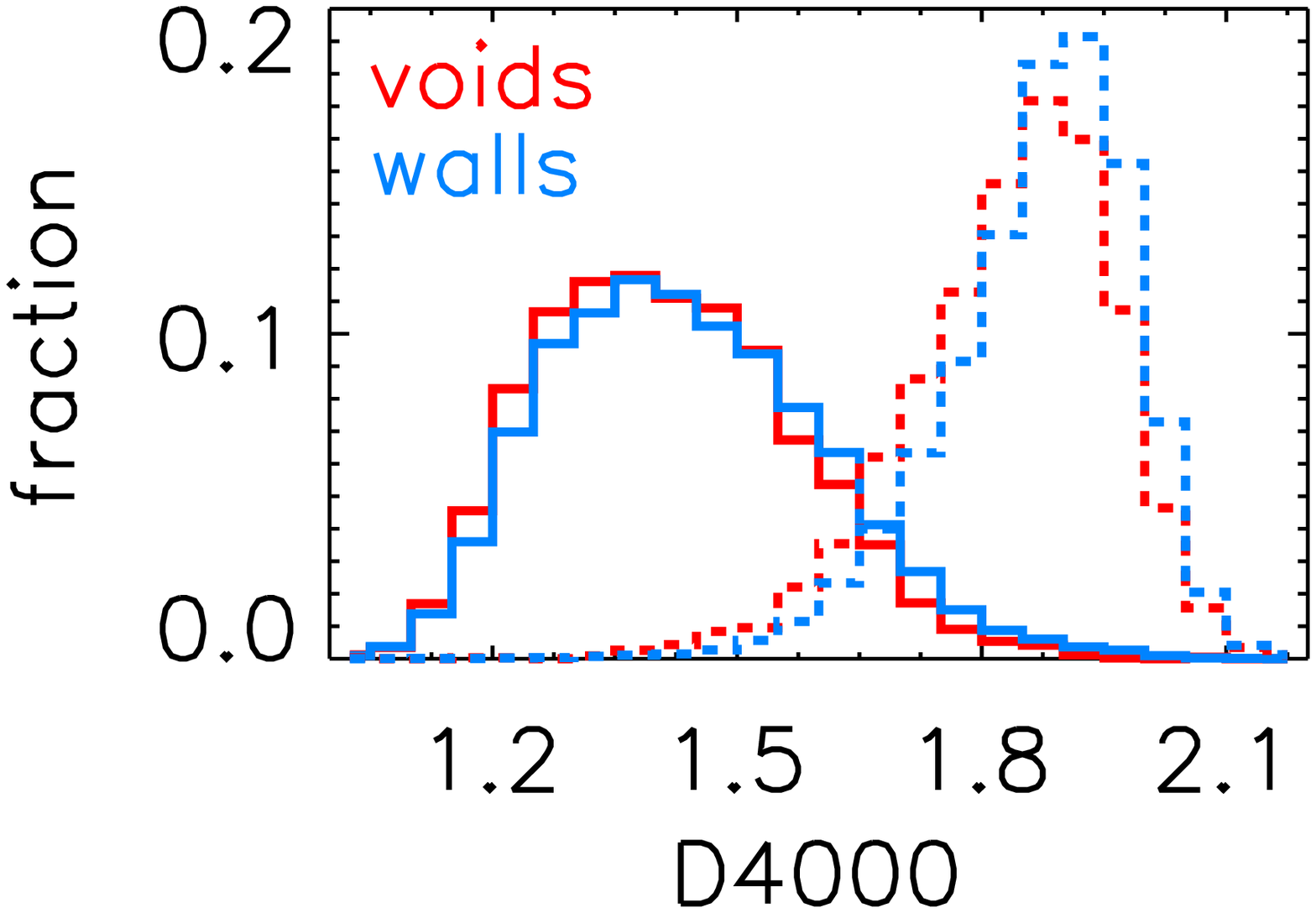}}\\
\subfigure[faint sample]{
\includegraphics[width=0.8\textwidth]{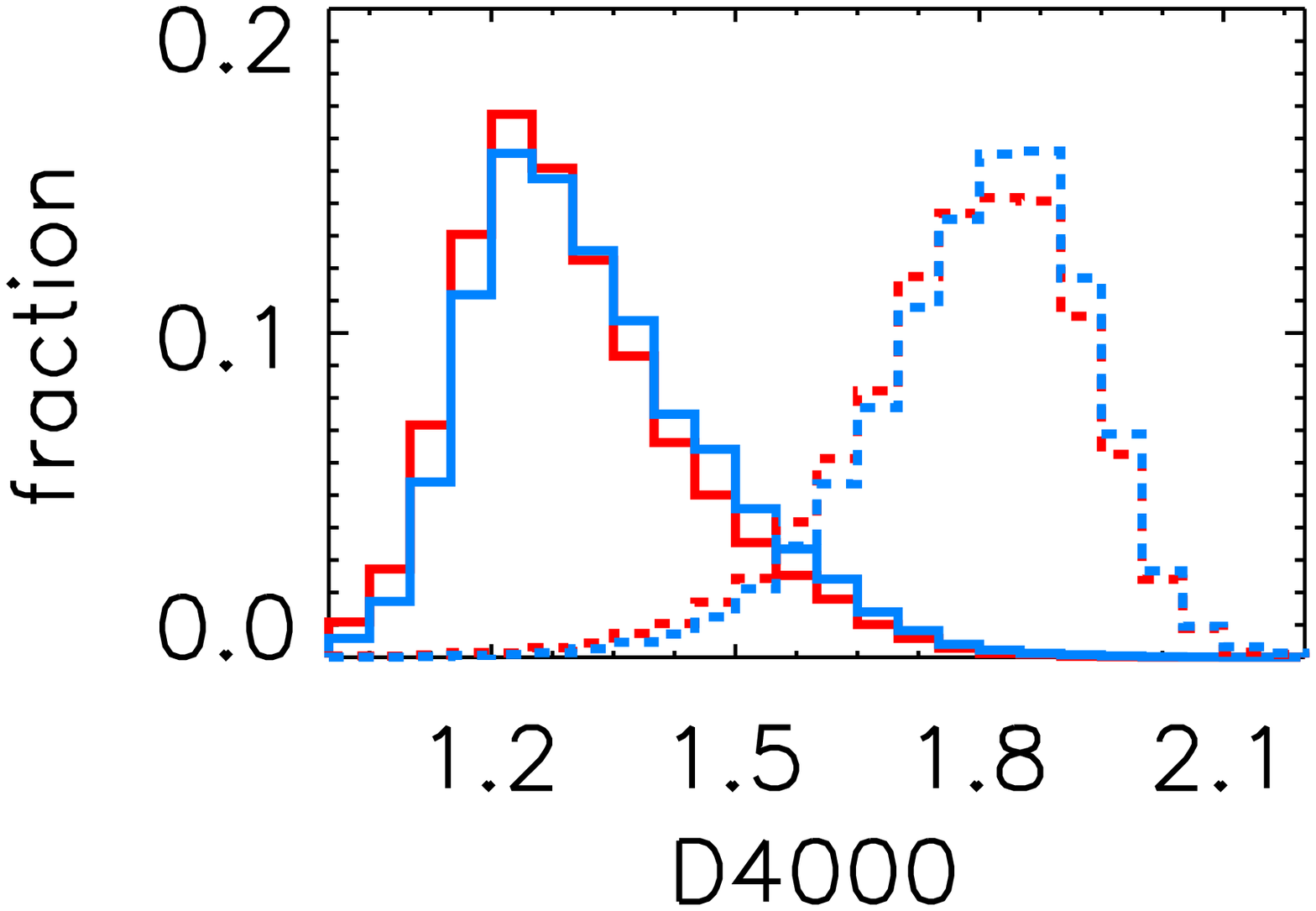}}\\
\subfigure[volume-limited sample]{
\includegraphics[width=0.8\textwidth]{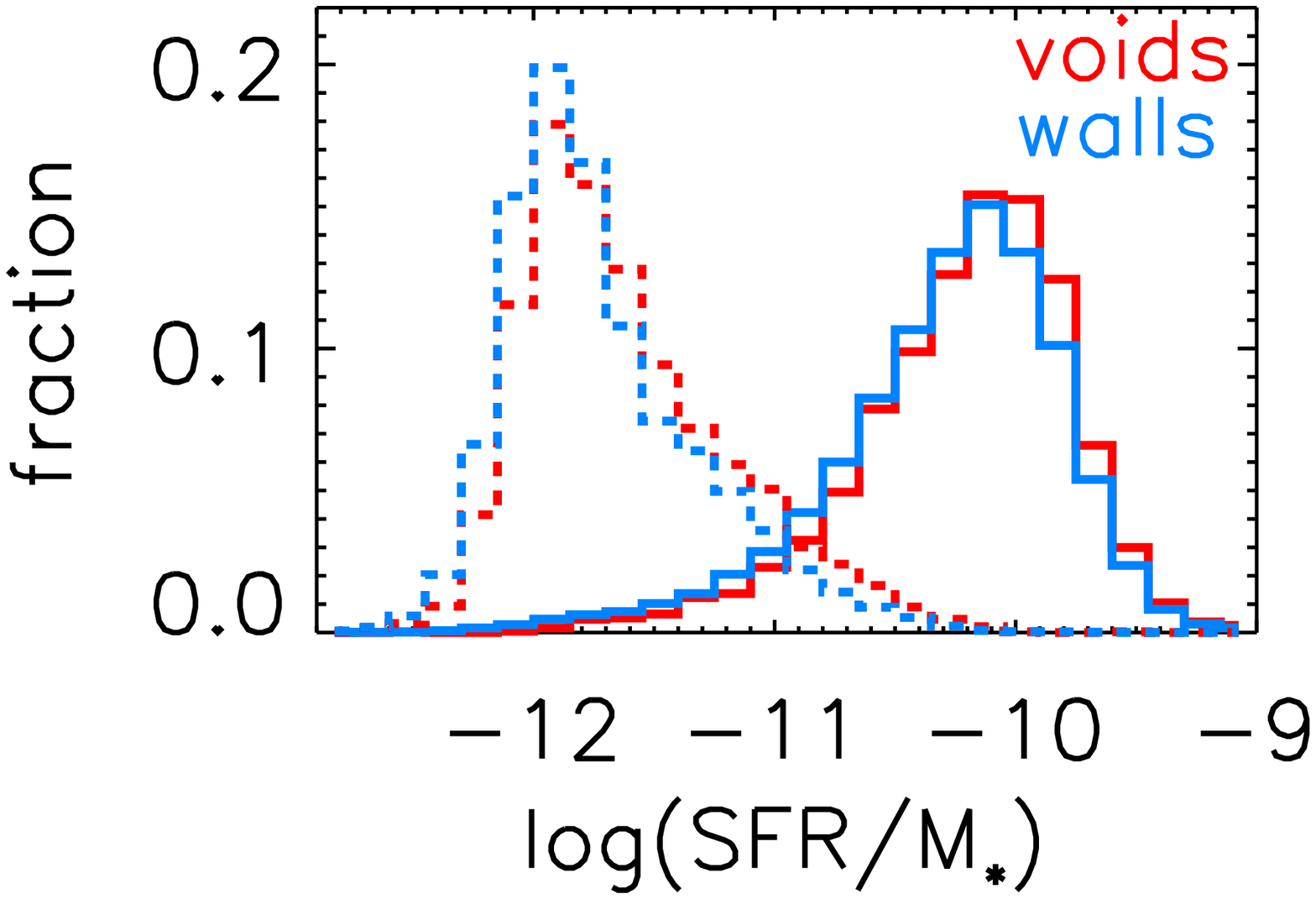}}\\
\subfigure[faint sample]{
\includegraphics[width=0.8\textwidth]{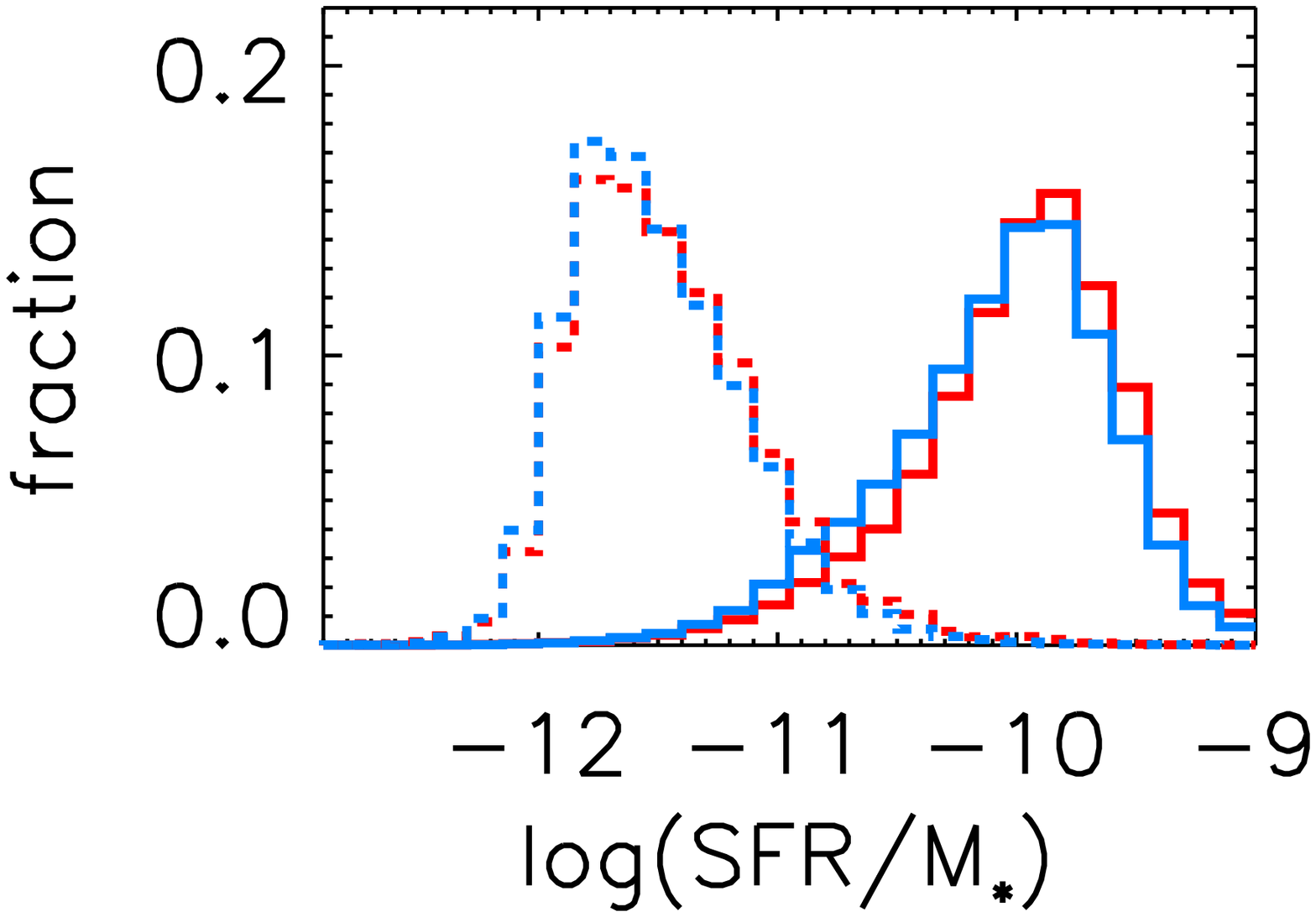}}
\caption{4000 \AA\  break (D4000) distributions and specific star formation rate (sSFR) distributions for both emission-line galaxies (solid) and non-emission-line galaxies (dashed) in voids (red) and in walls (blue) for the volume-limited sample and the faint sample.
\label{D4000_histo_full}}
\end{figure}

\begin{figure} 
\centering
\subfigure[volume-limited sample]{
\includegraphics[width=\textwidth]{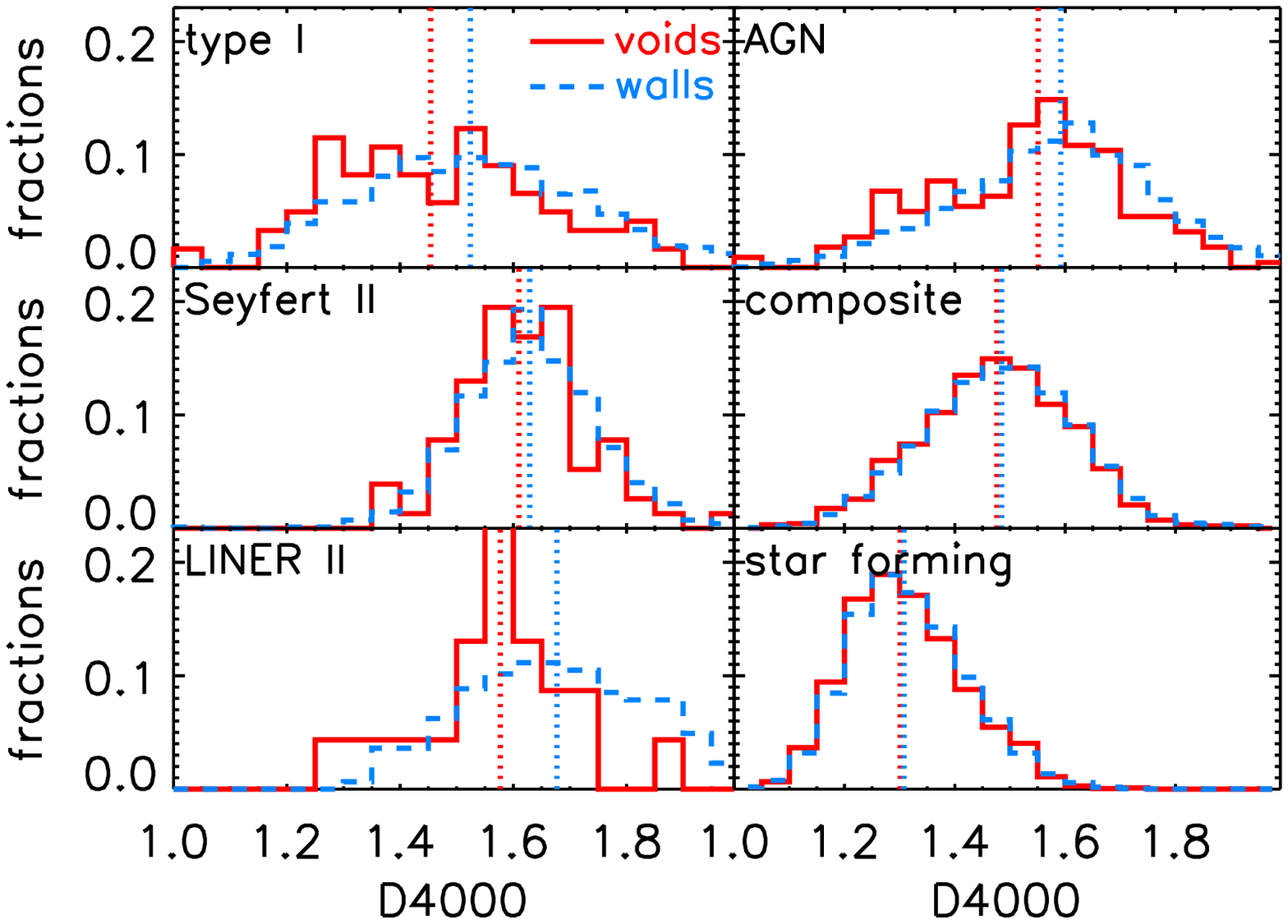}}\\
\subfigure[faint sample]{
\includegraphics[width=\textwidth]{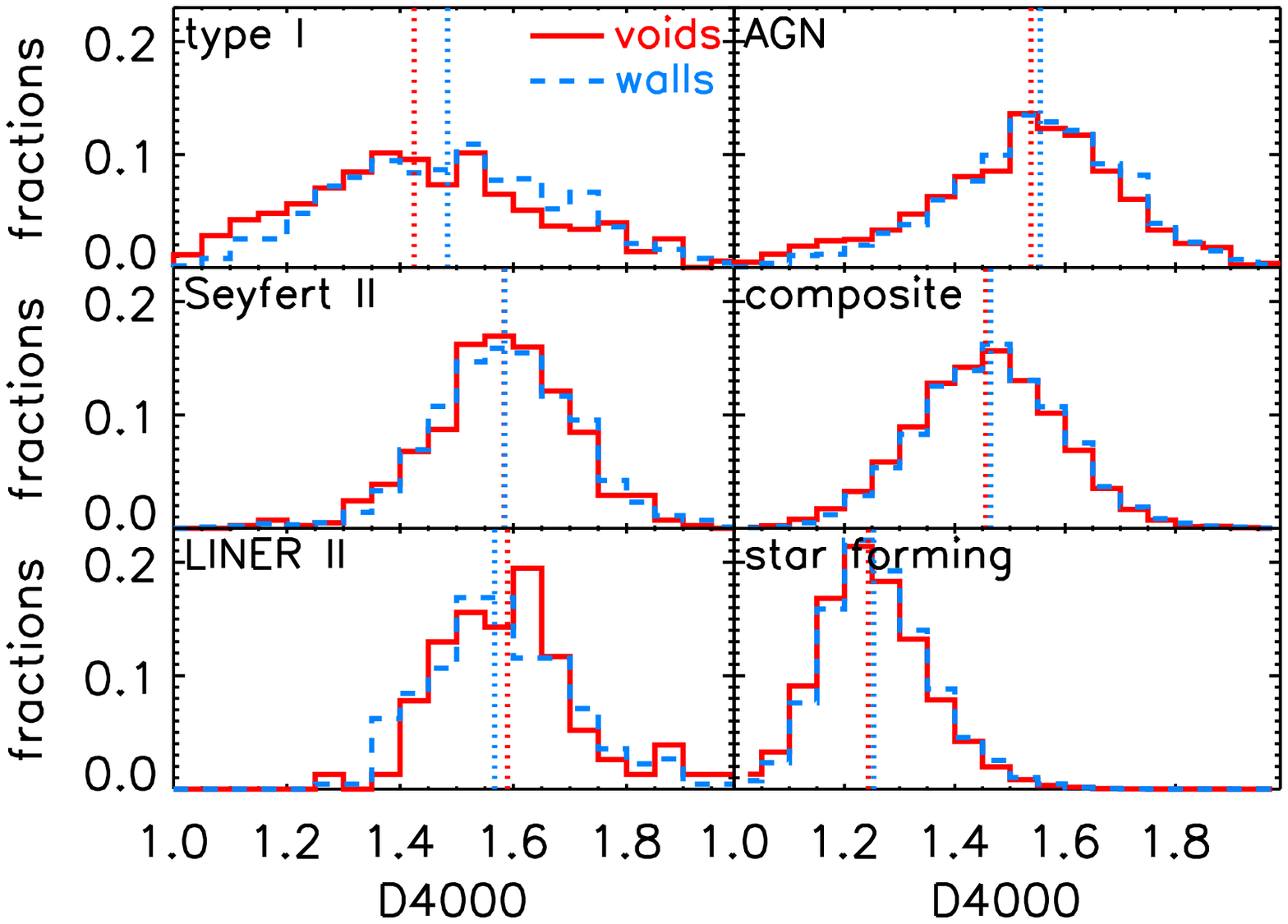}}
\caption{4000 \AA\  break (D4000) distributions for individual spectral type of galaxies in voids (red) and in walls (blue) for the volume-limited sample (up) and the faint sample (down). Note: AGNs = type Is + Seyfert IIs + LINER IIs. 
\label{D4000_distr}}
\end{figure}

\begin{figure} 
\centering
\subfigure[volume-limited sample]{
\includegraphics[width=\textwidth]{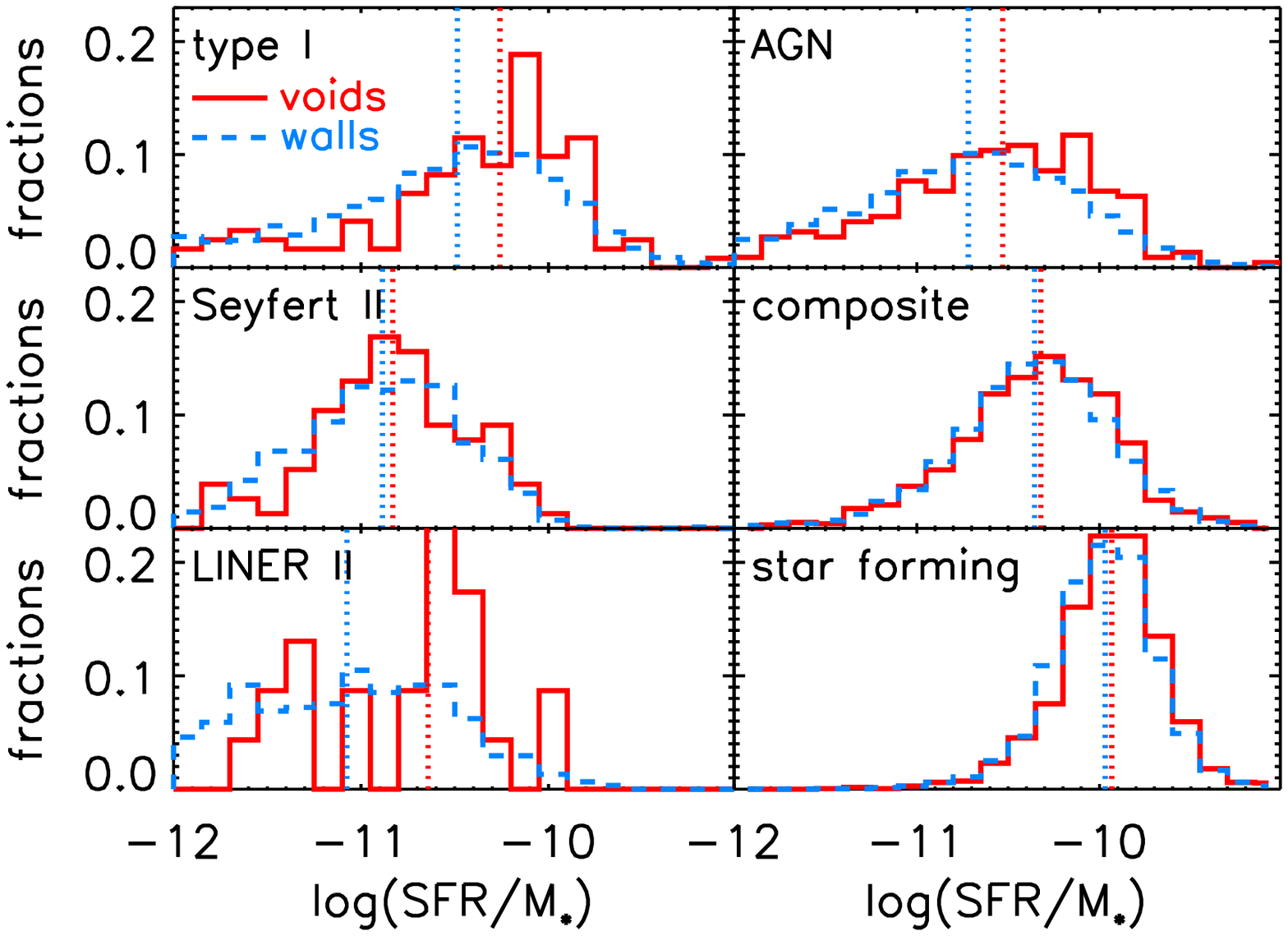}}\\
\subfigure[faint sample]{
\includegraphics[width=\textwidth]{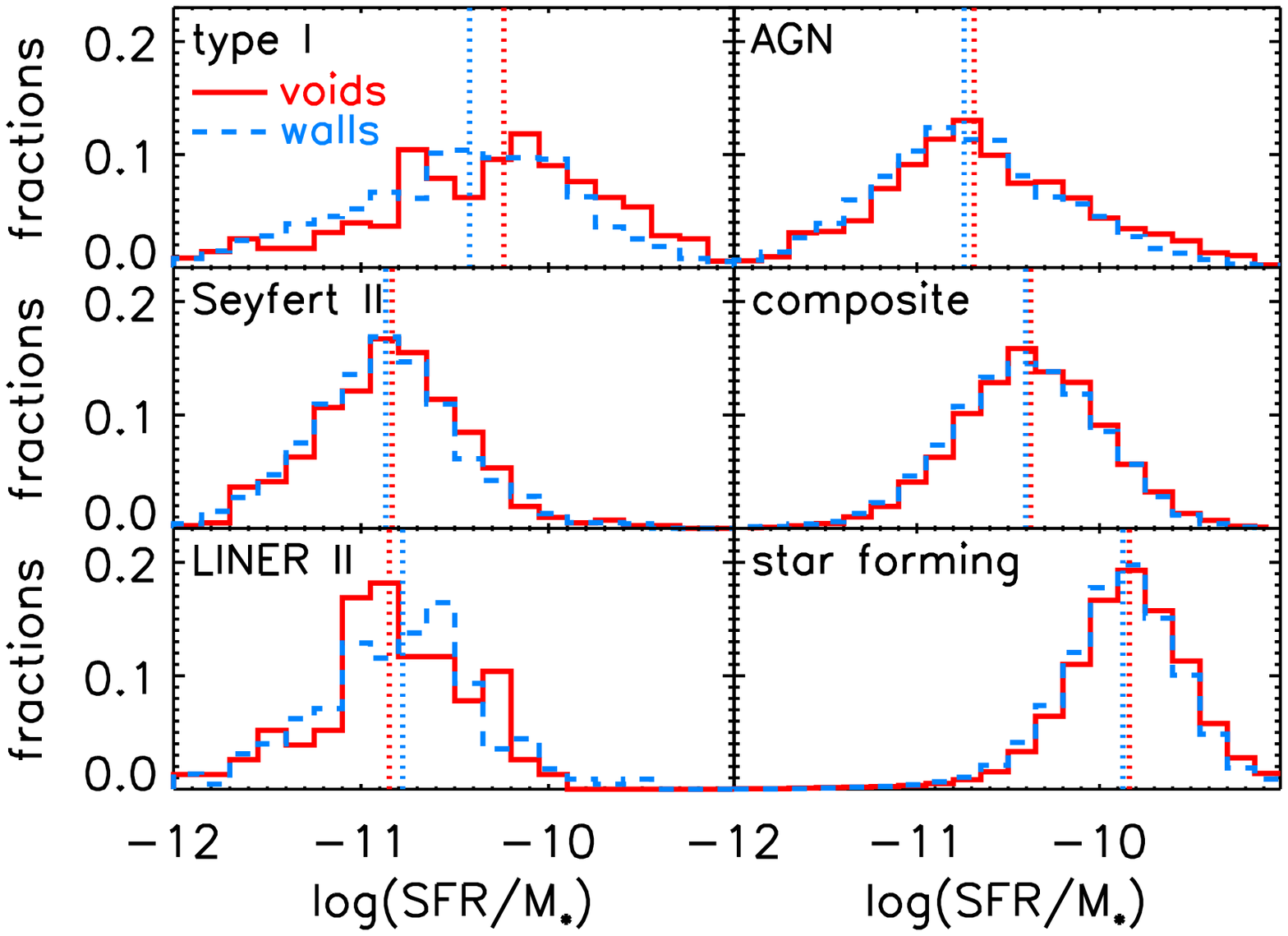}}
\caption{Specific star formation rate (sSFR) distributions for individual spectral type of galaxies in voids (red) and in walls (blue) for the volume-limited sample (up) and the faint sample (down).
Note: AGNs = type Is + Seyfert IIs + LINER IIs.
\label{sSFR_distr}}
\end{figure}

We use the 4000 \AA\ break (D4000) as an indicator of the stellar population. The 4000 \AA\ break is created by absorption lines located around the rest wavelength of 4000 \AA. The main contribution to the opacity comes from ionized metals. In hot stars, the elements are multiply ionized and the opacity decreases, so the 4000 \AA\ break will be small for young stellar populations and large for old metal rich galaxies. D4000 increases monotonically with time \citep{ka03stm,ka04}. 

Specific star formation rate (sSFR) is defined as the star formation rate normalized by the stellar mass ($\rm sSFR = SFR/M_* $). By studying the specific star formation rate difference between void galaxies and wall galaxies, we may understand the void environmental effect on the stellar mass assembly directly regardless the stellar masses.   

Figure \ref{D4000_histo_full} presents the D4000 distributions and sSFR distributions for void galaxies and the wall galaxies. We see that void galaxies are slightly younger and have higher specific star formation rates than wall galaxies. This result agrees with several previous studies. \cite{rojas05} used various indicators to study the star formation rates of void galaxies and found that void galaxies are of higher star formation rates than their wallcounter parts. \cite{ricc14} studied their $\sim 6000$ faint void galaxies and found that void galaxies have higher specific star formation rates than those in walls. \cite{hoyle12} adopt the exact same void galaxy catalog as our work and studied their photometric properties. They found that void galaxies are bluer than wall galaxies. We also note here that, similar with our results for stellar mass and luminosity analysis, the stellar population difference and star formation rate difference between void galaxies and the wall galaxies are very minor.

We further study the distributions of D4000 and the specific star formation rates for detailed classifications in Figure \ref{D4000_distr} and \ref{sSFR_distr} respectively. Each spectral type (type Is, Seyfert IIs, LINER IIs, composites, and star-forming galaxies) is of similar stellar populations and sSFRs in voids and in the walls, confirmed by the k-s test in Table \ref{kstests}.

Star formations are more intensively taking place in star-forming galaxies, as can be seen in Figure \ref{D4000_distr}: star-forming galaxies are of more newly formed stars and younger stellar populations. The object statistics (Section \ref{subsec_statistics}) shows that the fraction of star-forming galaxies increases dramatically from wall regions to underdense regions. Therefore, when considering AGNs, composites, and star-forming galaxies together, void emission-line galaxies are of younger stellar populations compared to emission-line galaxies in walls (Figure \ref{D4000_histo_full}). The younger stellar population of emission-line galaxies in void regions is totally caused by the number excess of star-forming galaxies in less dense regions (Section \ref{subsec_statistics}: Table \ref{classification}).

The same thing goes with the sSFR study. The star formation rates of star-forming galaxies is higher than that of any other spectral type (Figure \ref{sSFR_distr} and Table \ref{kstests}). Thus, the low shift of star formation rates of void emission-line galaxies, when mixing AGNs, composites, and star-forming galaxies together as shown in Figure \ref{D4000_histo_full}, is totally caused by the fact that star-forming galaxies are more abundant in cosmic voids (Section \ref{subsec_statistics}: Table \ref{classification}).

Our results on the sSFR of void galaxies agree with \cite{ricc14}. They used the specific star formation rate versus stellar mass diagram to separate star-forming galaxies from the rest of the galaxies (Section \ref{subsec_statistics}). They found that there are more star-forming galaxies in void regions and void galaxies have higher sSFR. They also found that the average sSFR of their star-forming galaxies and quiescent galaxies does not depend on the large-scale environment. This agrees with our results. 

\begin{figure} 
\centering
\includegraphics[width=\textwidth]{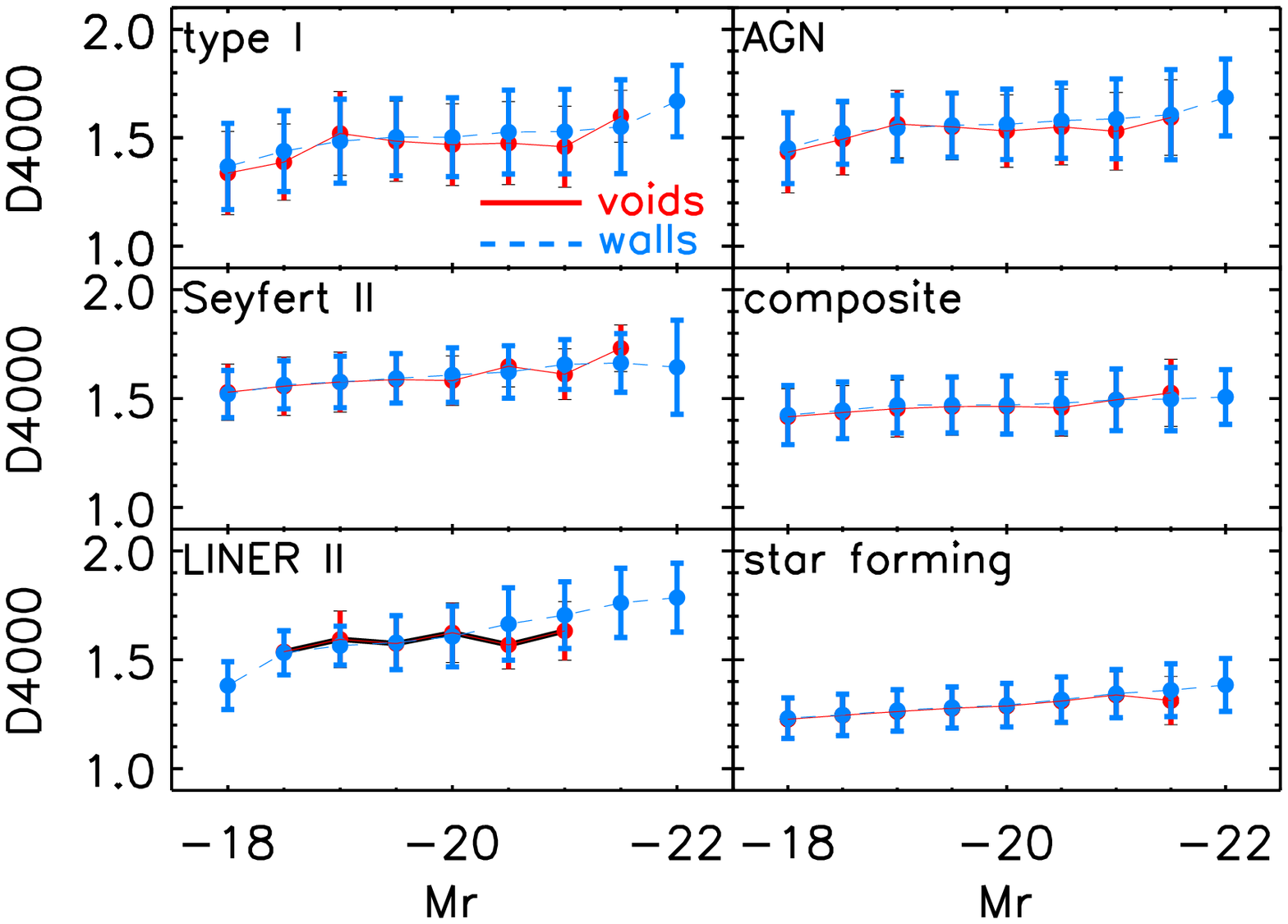}\\
\includegraphics[width=\textwidth]{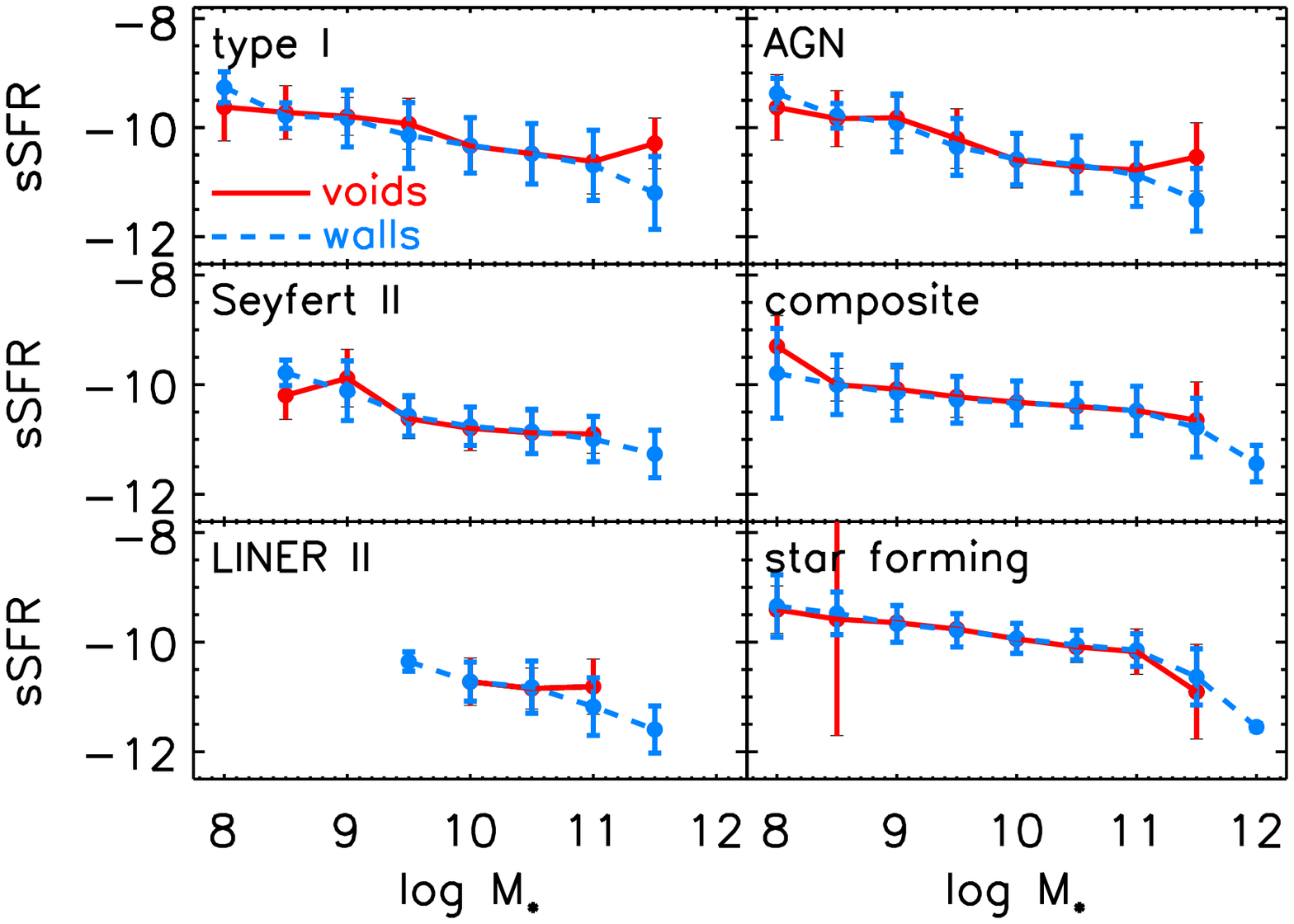}
\caption{Up: Mean values of the 4000 \AA\ break (D4000) as a function of $M_r$. 
Down: Mean values of sSFR as a function of $M_*$. 
Errors are standard deviation of galaxies in each bin. Data points are presented only for bins that include at least two objects.
\label{D-Mr_SFR-M*}}
\end{figure}
We explored whether large scale structures drive a primary correlation between star formation rate/stellar population and the galaxy density by studying D4000 as a function of $M_r$ and sSFR as a function of stellar mass in Figure \ref{D-Mr_SFR-M*}. Figure \ref{D-Mr_SFR-M*} shows us that at a fixed host brightness/stellar mass, there is no significant difference in stellar population and no significant difference in star formation rates between void galaxies and wall galaxies for all spectral types. This suggests that environment does not drive the primary correlation between star formation rate/stellar population and galaxy density for each type of galaxy. The results that void galaxies, when considering all spectral types together, have slightly younger stellar populations and higher specific star formation rates are purely resulted by the stellar mass difference between void galaxies and wall galaxies.

\subsection{Spectral Properties of Void AGNs}
\label{subsec_agn}

\subsubsection{AGN Fraction and AGN Luminosity}
\label{subsubsec_agn_lumi}

Figure \ref{bpt} and Table \ref{classification} show that AGNs do exist in voids. This is also found by \cite{co08}. 
The statistical results for AGNs have been presented in the last paragraph of Section \ref{subsec_statistics}.
The abundance of AGNs in voids and walls are similar within the sample uncertainties. Whether LINER IIs are included as AGNs or not does not affect this result.


\begin{figure} 
\centering
\subfigure[volume-limited sample]{
\includegraphics[width=\textwidth]{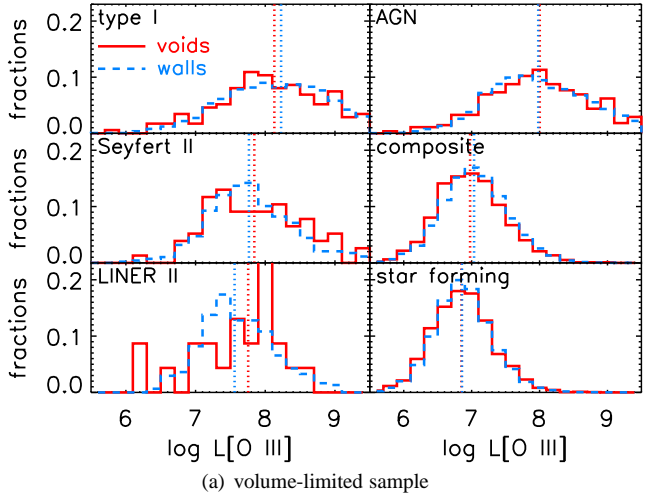}}\\
\subfigure[faint sample]{
\includegraphics[width=\textwidth]{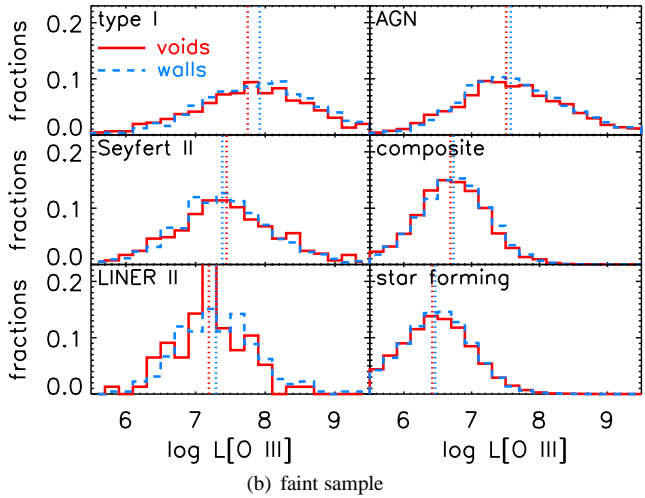}}
\caption{Up: Distributions of [O III] luminosity of each spectral type of galaxies for volume-limited void galaxies (red) and their wall counterparts (blue). Dotted lines are median values of each sub sample, red for void galaxies and blue for wall galaxies. 
The luminosities are extinction corrected as discussed in Section \ref{subsubsec_mpa}. Bottom: Same plot with the upper panel, but for faint sample. 
Note: AGNs = type Is + Seyfert IIs + LINER IIs.
\label{f.lo3}}
\end{figure}

We use the extinction-corrected [$\rm O\ _{III}$] luminosity ($L_{[O\ III]}$) \citep{heck04} to explore the luminosities of AGNs.  Figure \ref{f.lo3} shows the extinction-corrected $L_{[O\ III]}$ distribution for each spectral class. Our AGNs are not very luminous, with [$\rm O\ _{III}$] luminosities ranging from $10^6\ L_{\sun}$ to $10^9\ L_{\sun}$. We can see that for all three AGN classes (type I, Seyfert II, and LINER II) and composites, there is no significant difference in $L_{[O\ III]}$ between void AGNs and wall AGNs. This is confirmed by the k-s test results (Table \ref{kstests}).

\begin{figure} 
\centering
\subfigure[volume-limited sample]{
\includegraphics[width=0.9\textwidth]{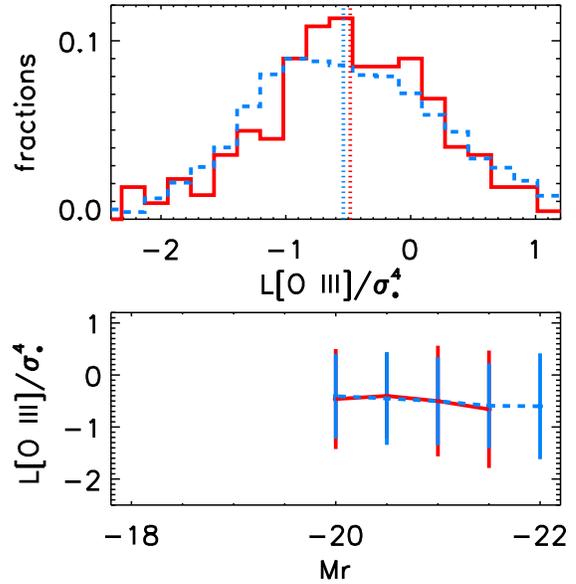}}\\
\subfigure[faint sample]{
\includegraphics[width=0.9\textwidth]{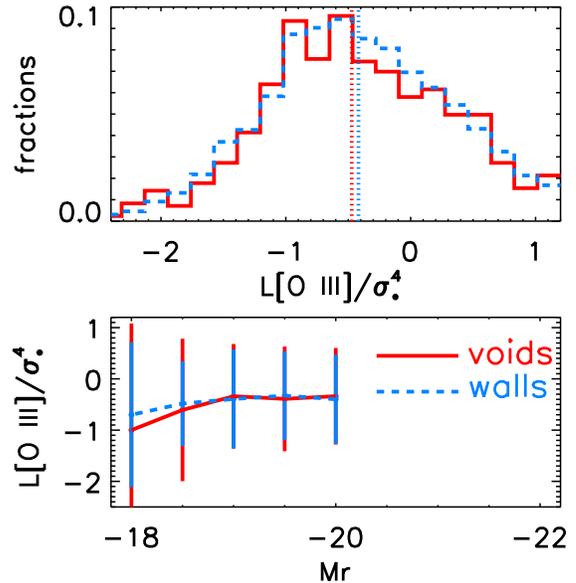}}
\caption{Distributions of $\rm L[O III]/\sigma_*^4$ for void AGNs (red) and wall AGNs (blue) and mean values of $\rm log\ L_{[O_{\ III}]}/\sigma_*^4$ as a function of $\rm M_r$, for galaxies in voids (red) and for galaxies in walls (blue) for the volume-limited sample and the faint sample. The error bars represent the standard deviation of the mean. The stellar velocity dispersions ($\sigma_*$) are in $\rm km\ s^{-1}$. The luminosities are extinction corrected as discussed in Section \ref{subsubsec_mpa}. Data points are presented only for bins that include at least two objects.
\label{f.edd}}
\end{figure}

The stellar velocity dispersion can be used as an estimate of the black hole mass \citep{Gebhardt2000,Ferrarese2000,tremaine02,Kormendy_Ho2013}. Thus, the quantity $L_{[O\ III]}/\sigma_*^4$ is proportional to $ \rm \Gamma =  L_{bol}/L_{Edd} $. Figure \ref{f.edd} indicates that the $L_{[O\ III]}/\sigma_*^4$ of AGNs also do not change significantly with large scale environments.

\subsubsection{Type I vs. Type II}
\label{subsubsec_IvsII}

The AGN unification model \citep{antonucci93} hypothesizes that both type Is and type IIs are intrinsically the same, seen from different inclinations. Type II AGNs are viewed edge-on, so the central accretion disc and the broad-line region are obscured by the dusty torus, while type I AGNs are viewed face-on. Thus, according to the unification scheme of AGNs, both types are expected in voids and in walls, and the number ratio ($f_{12}$) of type Is vs. type IIs should not depend on environment. 

\begin{figure} 
\centering
\includegraphics[width=0.98\textwidth]{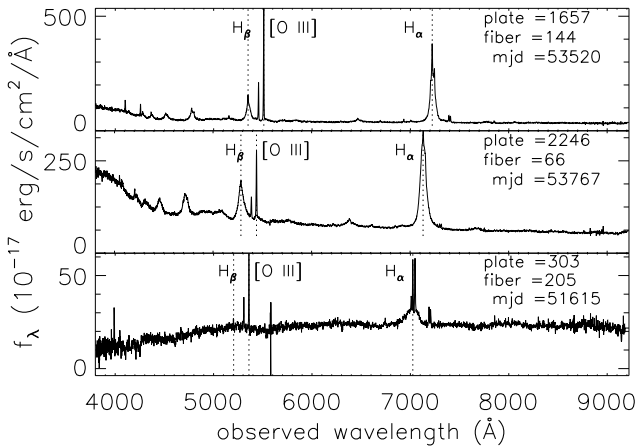}
\caption{Three examples of our void type I AGNs. Top two panels show two examples of quasar-like type I AGN spectra. 
The bottom panel is one example of type I AGNs with both broad $H_\alpha$ component and narrow $H_\alpha$ component.
Dashed lines presents the positions of $H_\alpha$, [O III], and $H_\beta$ lines in the observed frame.
\label{f.type1}}
\end{figure}

Our study showed  for the first time that type Is exist in void regions. Type I AGNs are identified as the emission-line galaxies with a FWHM of the $H_\alpha$ emission greater than 1200 km/s. Figure \ref{f.type1} gives three examples of our void type I 
AGNs. Some show quasar-like spectra, while some show both broad and narrow $H_\alpha$ emissions.

We study $f_{12}$ based on the statistics shown in Table \ref{classification}. This ratio does not change significantly from the underdense environment to the more crowded environment. $f_{12}$ are approximately 0.4 in the magnitude-limited sample, 0.5 in the volume-limited sample, and 0.3 in the faint sample. It looks that the $f_{12}$ depends weakly on large scale environments, which agrees with the unification model. 

Hosts of type Is and type IIs are similarly luminous and comparably massive (Figure \ref{Mr_distr} and Figure \ref{stmass_distr}). Figure \ref{f.lo3} shows us that type I AGNs are of higher bolometric luminosities than type IIs. This is seen in both void and wall regions.
\subsection{The Small Scale Environment: Nearest Neighbor Statistics}
\label{subsec_small_scale}

In addition to the large-scale environment, which is probed by void and wall regions, in this section we would also like to investigate whether the small-scale environment affect the spectral properties of galaxies.

We assess the local environment of galaxies by looking at the nearest neighbor
distances of the galaxies in our sample.  The nearest neighbor distance (nn)
is determined by finding the distance to the nearest galaxy in the volume
limited galaxy catalog (M$_{r}$ < -20.09).  By using the volume-limited galaxy
catalog as the basis set of structure, we maintain the redshift fidelity of
the sample.  

\begin{figure}
\centering
\includegraphics[width=\textwidth]{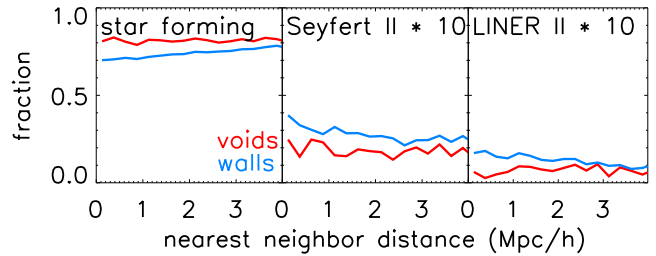}
\caption{The fraction of emission-line galaxies that are identified as star-forming galaxies, Seyfert IIs, or LINER IIs as a function of nearest neighbor distance in voids (red) and in walls (blue). The fraction of Seyfert IIs and LINER IIs are multiplied by 10 for presentation purpose.}
\label{nnbydist}
\end{figure}

Figure \ref{nnbydist} shows the fraction of emission-line galaxies in both the
wall and void samples that are identified as star-forming galaxies, Seyfert IIs, or LINER IIs.
The flatness of the distribution shows that at any given nearest neighbor
distance, there is no apparent change in the distribution of identified galaxy
types.  This hints at the overall null effect of the small scale structure
towards the distribution.  What can be seen is the preference of emission line
galaxies detected in voids to be star-forming galaxies, this is observed
by the vertical separation of the two galaxy samples in the figures.

\begin{figure}
\centering
\includegraphics[width=\textwidth]{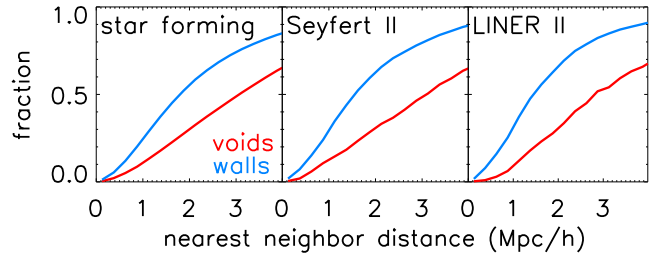}
\caption{The cumulative fraction of individual galaxy types that have nearest neighbors at the give distance for void regions (red) and wall regions (blue).}
\label{nnbytype}
\end{figure}

Figure \ref{nnbytype} shows the fraction of individual galaxy types that have
nearest neighbors at the given distance.  In each panel, a trend can be seen
where the wall sample has higher percentages of galaxies with nearest
neighbors that are closer compared to the void sample.  The distribution is
almost identical amongst all three galaxy types.  The low number of galaxies
in the Seyfert II and LINER II samples contribute to the non-smooth distribution
seen, but the overall trend is similar.

We also examine the possible effect of nearest neighbors on the galaxy type as
a function of the absolute magnitude of the galaxy.  The distribution of
galaxy types as a function of absolute magnitude was previously discussed in
section 3.2.1.

\begin{figure}
\centering
\includegraphics[width=\textwidth]{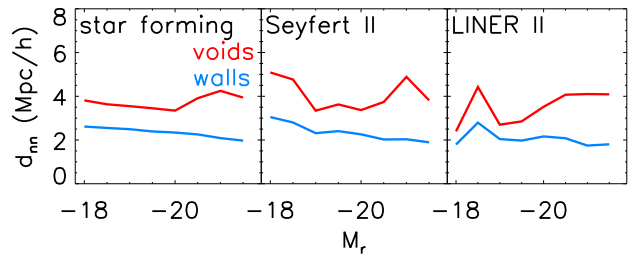}
\caption{The nearest neighbor distance ($d_{nn}$) as a function of r-band absolute magnitude for different spectral type of galaxies in voids (red) and in the walls (blue).}
\label{nndistance}
\end{figure}

We see in Figure \ref{nndistance} that the nearest neighbors of void galaxies
are systematically further away than the wall counterparts
equivalently in all three galaxy types discussed.  There also appears to be an
increase in the nearest neighbor distance as a function of r band absolute
magnitude. 

\section{Discussions}
\label{sec_discuss}

\subsection{Environmental Effects On The Stellar Mass Assembly of Galaxies}
\label{subsec_dis_gal}

 There are many previous works studying the relationship between galaxy properties and large scale environments. People found that void galaxies are less massive, less luminous, bluer, of younger stellar populations, have higher specific star formation rates, and later morphology types \citep[e.g.,][]{LF,MF,rojas05,hoyle12}. However, it was not clear how environments influence these properties of galaxies. It could be that the underdense universe make individual spectral type of galaxies, such as AGNs, composites, and star-forming galaxies different compared to their wall counterparts, or it could be that in void regions, individual spectral type of galaxies have similar properties with their wall counterparts, but the partition of void galaxies is different from that of wall galaxies, for example, void galaxies may have a higher fraction of one certain type of galaxies. Our study clearly demonstrate that it is the second case.

 In our study, for each individual spectral type, there is no significant difference in spectral properties (stellar mass, luminosity, stellar population, and star formation rate) between void galaxies and the wall galaxies  (Figure \ref{Mr_distr}, \ref{stmass_distr}, \ref{D4000_distr}, and \ref{sSFR_distr}; Table \ref{kstests}). However, the partition of void galaxies and wall galaxies are significantly different. There are 30\% more star-forming galaxies in void regions (Table \ref{classification}). Since star-forming galaxies are less luminous, less massive, of younger stellar populations, and have higher specific star formation rates (Figure \ref{Mr_distr}, \ref{stmass_distr}, \ref{D4000_distr}, and \ref{sSFR_distr}; Table \ref{kstests}), when considering all spectral types of galaxies together, void galaxies are slightly fainter, less massive, younger, and of higher star formation rates than field galaxies. 

Why would there be more star-forming galaxies in voids? In Figure\ref{D-Mr_SFR-M*}, we find that the relation between the spectral properties of galaxies and their stellar masses do not vary with galaxy density. This suggests that environment does not drive the primary correlation between star formation rate and stellar mass.A possible explanation for more star-forming galaxies in voids could be that underdense environments have less massive halos \citep{ben03}, thus form less massive galaxies (section 3.2.1). According to the color magnitude diagram \citep{bell04} of galaxies, less massive galaxies are more likely to locate in the blue cloud and be identified as star-forming galaxies. Therefore, star-forming galaxies are more abundant in void regions than in the fields, making the overall void galaxy spectral properties shift a little bit to the spectral properties of star-forming galaxies.  In summary, we think that void environments have less massive halos, thus more galaxies with smaller stellar masses. Given a certain stellar mass, all other spectral properties of galaxies are similar for individual types of void galaxies and their wall counterparts. The void environmental effect is shifting the mass function of galaxies and halos to the low mass end slightly \citep{MF,LF}, resulting in the high fraction of star-forming galaxies in voids. 

\subsection{Environmental Effects On The Growth of Black Holes}
\label{subsec_dis_BH}

Our study shows that there is no significant environmental effects on the growth of black holes in void regions. The fractions of AGNs in void regions and in wall regions are similar.  Other AGN properties, such as accretion rates (estimated by $\rm L[O_{\ III}]$) and Eddington ratios (estimated by $\rm L[O_{\ III}]/\sigma^4_*$), do not vary with galaxy densities (Section \ref{subsubsec_agn_lumi}). 

Major merger can also be a possible trigger of AGNs. Tidal torques excited during a gas rich merger can lead to rapid inflows of gas into the centers of galaxies and feed rapid black hole growth, as supported by several hydrodynamic numerical simulations \citep{hernquist89,barnes91,barnes96,dimatteo05,hopkins06,debuhr11} and observations \citep{sanders88}. We explored the small scale environment for void galaxies in Section \ref{subsec_small_scale} and found that all spectral types of void galaxies are less clustered than their counterparts in walls Figure \ref{nnbytype} and \ref{nndistance}, indicating that void AGNs are less likely to be involved in major merger events than wall AGNs. The AGN fractions in our void galaxy sample and in our wall galaxy sample are comparable. Our AGNs are low luminosity AGNs with ($L_{[O\ III]} \sim 10^8\ L_{\sun}$) . Therefore, the major merger is unlikely to be the dominant triggering mechanism for our low luminosity AGNs.

Thus, AGNs in our study should be mainly triggered by other mechanisms than major merger or large-scale environments. Secular evolution is one possible mechanism. Galactic bars can also efficiently transport gas from the outer disk to the central kiloparsec scale \citep{atha03,jogee06} by reducing angular momentum, as demonstrated by a number of numerical simulations \citep[e.g.,][]{shen04,atha05,bour05,hopkins10}, even though there is still a discrepancy in observational studies on whether or not bars trigger AGN activity \citep{ars89,knap00,laine02,moles95,mcl95,mu97,ho97,laur04,hao09,lee12}. Since we did not explore morphologies for our galaxy sample, the secular evolution can not be ruled out as a main trigger of AGN activity for our low luminosity AGNs ($L_{[O\ III]} \sim 10^8\ L_{\sun}$).

\subsection{Expected Results in Deeper Surveys}
\label{deeper_surveys}

It is suggested that void regions can contain more fainter galaxies in deeper surveys \citep{Weygaert1993,gottl2003,Alpaslan2014}. These fainter galaxies can form their structures. For example, \cite{Alpaslan2014} studied the Galaxy And Mass Assembly survey (GAMA), which is 2-3 magnitudes fainter than the main SDSS survey. They found fine structures -- tendrils which is coherent thin chains of galaxies. The tendril structures on average contain six galaxies and span 10 Mpc. \cite{Alpaslan2014} further found that 25\% of the SDSS void galaxies from \cite{pan12} are isolated galaxies, 64\% form fine tendrils (the rest 11\% are actually in filaments). If we consider the large-scale structures (for example, \cite{pan12} only consider regions larger than 10 Mpc as a void region), these tendril galaxies are typically still considered as void galaxies.  If a consistent definition of voids is kept, fainter surveys will find similar void galaxies as SDSS.

For a deeper survey, we expect the overall luminosities for all of the wall galaxies and void galaxies to be slightly lower, since deeper surveys can detect fainter galaxies. However, we speculate that there is no significant difference in the luminosities between the void and wall galaxies. As seen in Table \ref{classification}, the AGN fraction of the faint sample is lower than that of the volume limited sample. If we extend this trend, we may find that AGN fraction decreases as galaxies get less luminous. Thus, in a deeper survey, the overall AGN fraction may be a little bit lower compared to what we presented in Section \ref{subsec_agn}, but the comparison of the AGN fraction between void and wall galaxies may still be similar. We do not yet have the data to check further if the spectral properties of the ``isolated'' galaxies and ``tendril'' galaxies are the same, we leave it for future investigations.

\section{conclusions}
\label{sec_conclusions}

We have analyzed the object statistics, luminosities, stellar masses, stellar populations, specific star formation rates, and AGN properties of 75,939 void galaxies selected from SDSS DR7. We find that:\\

\indent (i) 
There are more emission-line galaxies in voids than in walls, with a percentage difference greater than 30\%. The high fraction of emission-line galaxies in void regions is accounted by the rich abundance of star-forming galaxies in voids: There are $\sim$ 30\% more star-forming galaxies in underdense regions compared to that in walls. The fractions of other types of galaxies, such as AGNs and composites, do not diverge too much from voids to walls.\\

\indent(ii) Void galaxies of individual spectral types, such as AGNs, composites, and star-forming galaxies, are similar in all kinds of spectral properties (luminosities, stellar masses, stellar populations, and specific star formation rates) compared to their wall counterparts. However, since void galaxies have $\sim$ 30\% more star-forming galaxies, and star-forming galaxies are less luminous, less massive, younger in stellar populations and higher in star formation rates than other types of galaxies (Figure \ref{Mr_distr}, \ref{stmass_distr}, \ref{D4000_distr}, \ref{sSFR_distr}; Table \ref{kstests}), void galaxies, when considering all types of galaxies together, show minor different properties: They are slightly less luminous, less massive, of younger stellar populations, and have higher specific star formation rates than wall galaxies. All the minor differences are totally caused by the high fraction of star-forming galaxies in void regions.\\

\indent (iii) We confirm that AGNs do exist in voids (Figure \ref{bpt}, Table\ref{classification}) \citep{co08}. We also note that type I AGNs are for the first time found in void regions in this work. AGNs in voids are similarily abundant as in walls (Table \ref{classification}). Our small scale study of environments of galaxies (Section \ref{subsec_small_scale}) shows that void AGNs are less clustered than their wall counterparts, thus less likely to be involved in galaxy-galaxy interactions. Therefore, major merger may not be the dominant triggering machanism of our optically selected AGNs. We note here that this conclusion should only be applied for AGNs with the luminosity range: $10^6\ L_{\sun} < L_{[O\ III]} < 10^9\ L_{\sun}$ that we probe in this paper.\\

\indent(iv)  The intrinsic [O III] luminosities, and the Eddington ratios of void AGNs are comparable to those of the wall AGNs (Figure \ref{f.lo3}, \ref{f.edd} and Table \ref{kstests}), implying that underdense environments do not affect the growth of black holes.

\begin{table}[t]

\caption{KS Test Results\label{kstests}}

\begin{tabular}{@{}c@{}}

\centering

   \begin{minipage}[t]{1.\textwidth}

     \begin{tabular*}{0.98\textwidth}[t]{cccccc}\hline

       \hline\hline
       
$\rm\bf log M_*/M_\sun$ \bf (dex) & in voids  & in walls &v-w\tablenotemark{a} &  D\tablenotemark{b}  &  PROB\tablenotemark{c}  \\\hline
\multicolumn{6}{c}{\bf volume limited sample} \\\hline
   type I              &   10.739   &   10.783  & -0.044  &  0.128 & 4.62e-2  \\
  Seyfert              &   10.778   &   10.816  & -0.038  &  0.134 & 1.39e-1  \\
   LINER               &   10.778   &   10.840  & -0.062  &  0.143 & 7.40e-1  \\\hline
    AGN                &   10.753   &   10.799  & -0.046  &  0.123 & 3.55e-3  \\
  composite            &   10.679   &   10.718  & -0.039  &  0.100 & 2.96e-8  \\
 star forming          &   10.438   &   10.477  & -0.039  &  0.068 & 4.04e-7  \\\hline
emission line galaxies &   10.569   &   10.626  & -0.057  &  0.093 & 2.36e-23  \\
  no emission          &   10.769   &   10.842  & -0.073  &  0.134 & 0.00e+0 \\\hline
     total             &   10.697   &   10.783  & -0.086  &  0.145 & 0.00e+0 \\\hline
\multicolumn{6}{c}{\bf faint sample} \\\hline
   type I               &   10.232   &   10.328   &   -0.096 &  0.156 & 1.27e-5  \\
  Seyfert               &   10.400   &   10.407   &   -0.007 &  0.033 & 9.03e-1  \\
   LINER                &   10.349   &   10.422   &   -0.073 &  0.171 & 6.17e-2  \\\hline
    AGN                 &   10.329   &   10.376    &  -0.047 &  0.089 & 1.67e-4  \\
  composite             &   10.298   &   10.322    &  -0.024 &  0.049 & 6.17e-8  \\
 star forming           &    9.700   &    9.781    &  -0.081 &  0.065 & 0.00e+0  \\\hline
emission line galaxies  &    9.757   &    9.865    &  -0.108 &  0.076 & 0.00e+0  \\
  no emission           &   10.295   &   10.322    &   0.027 &  0.039 & 1.68e-22  \\\hline
     total              &    9.957   &   10.100    &  -0.143 &  0.102 & 0.00e+0  \\\hline\hline\\\hline


\hline\hline
$\rm\bf lg L[O\ {III}]/L_\sun$ & in voids  & in walls &v-w\tablenotemark{a} &  D\tablenotemark{b}  &  PROB\tablenotemark{c}  \\\hline

\multicolumn{6}{c}{\bf volume limited sample} \\\hline
   type I              &   8.130   &   8.240  & -0.109  &  0.053 & 7.39e-1  \\
  Seyfert              &   7.843   &   7.766  &  0.077  &  0.090 & 5.88e-1  \\
   LINER               &   7.755   &   7.560  &  0.195  &  0.200 & 3.25e-1  \\\hline
    AGN                &   8.003   &   7.984  &  0.019  &  0.035 & 9.14e-1  \\
 composite             &   6.977   &   7.036  & -0.059  &  0.074 & 9.71e-5  \\
star forming           &   6.850   &   6.858  & -0.008  &  0.037 & 2.36e-2  \\\hline
emission line galaxies &   6.886   &   6.930  & -0.043  &  0.042 & 2.06e-6  \\\hline
\multicolumn{6}{c}{\bf faint sample} \\\hline
   type I              &   7.757   &   7.932  & -0.175  &  0.118 & 2.95e-4  \\
  Seyfert              &   7.445   &   7.383  &  0.062  &  0.059 & 2.56e-1  \\
   LINER               &   7.191   &   7.293  & -0.102  &  0.153 & 1.26e-1  \\\hline
    AGN                &   7.513   &   7.579  & -0.066  &  0.066 & 5.67e-3  \\
 composite             &   6.687   &   6.733  & -0.046  &  0.040 & 1.72e-5  \\
star forming           &   6.421   &   6.463  & -0.042  &  0.038 & 5.39e-29 \\\hline
emission line galaxies &   6.394   &   6.445  & -0.051  &  0.038 & 1.74e-43 \\\hline\hline\\\hline

\hline\hline
\bf D4000  & in voids  & in walls &v-w\tablenotemark{a} &  D\tablenotemark{b}  &  PROB\tablenotemark{c}  \\
 \hline
\multicolumn{6}{c}{\bf volume limited sample} \\\hline
   type I              &   1.454  &   1.524  & -0.070  &  0.156 & 7.31e-3  \\
  Seyfert              &   1.620  &   1.629  & -0.019  &  0.101 & 4.44e-1  \\
   LINER               &   1.577  &   1.677  & -0.100  &  0.391 & 1.92e-3  \\\hline
    AGN                &   1.551  &   1.592  & -0.041  &  0.141 & 4.92e-4  \\
  composite            &   1.476  &   1.485  & -0.009  &  0.034 & 2.38e-1  \\
 star forming          &   1.301  &   1.308  & -0.007  &  0.039 & 1.39e-2  \\\hline
emission line galaxies &   1.373  &   1.400  & -0.027  &  0.067 & 3.93e-12 \\
  no emission          &   1.806  &   1.863  & -0.057  &  0.143 & 0.00e+0  \\\hline
     total             &   1.591  &   1.761  & -0.170  &  0.162 & 0.00e+0  \\\hline
\multicolumn{6}{c}{\bf faint sample} \\\hline
   type I              &   1.425   &   1.484  & -0.059  &  0.129 & 5.99e-4  \\
  Seyfert              &   1.584   &   1.585  &  0.001  &  0.039 & 7.61e-1  \\
   LINER               &   1.590   &   1.567  &  0.023  &  0.110 & 4.70e-1  \\\hline
    AGN                &   1.538   &   1.555  & -0.017  &  0.072 & 3.77e-3  \\
  composite            &   1.456   &   1.465  & -0.009  &  0.034 & 4.80e-4  \\
 star forming          &   1.243   &   1.253  & -0.010  &  0.047 & 0.00e+0 \\\hline
emission line galaxies &   1.260   &   1.277  & -0.017  &  0.057 & 0.00e+0  \\
  no emission          &   1.707   &   1.748  & -0.041  &  0.073 & 0.00e+0 \\\hline
     total             &   1.325   &   1.405  & -0.080  &  0.129 & 0.00e+0 \\\hline\hline
\\

     \end{tabular*}

   \end{minipage}

\hfil

   \begin{minipage}[t]{1.\textwidth}

     \begin{tabular*}{0.98\textwidth}[t]{cccccc}\hline

       \hline\hline
        $\bf M_r$ \bf(mag) & in voids  & in walls &v-w\tablenotemark{a} &  D\tablenotemark{b}  &  PROB\tablenotemark{c}  \\\hline
\multicolumn{6}{c}{\bf magnitude limited sample} \\\hline
   type I              &   -19.846   &   -20.394  & 0.548  &  0.409 & 0.00e+0  \\
  Seyfert             &   -19.665   &   -20.074  & 0.409  &  0.338 & 9.85e-40  \\       
   LINER               &    -19.737  &   -20.213  & 0.476  &  0.351 & 1.15e-9  \\ \hline
    AGN                 &   -19.747  &   -20.268  & 0.521   &  0.383 & 0.00e+0  \\
  composite          &    -19.654  &   -20.044  & 0.390  &  0.320 & 0.00e+0  \\
 star forming        &   -19.094   &   -19.416  & 0.322  &  0.160 & 0.00e+0  \\\hline
emission line galaxies &  -19.158  &   -19.599  & 0.441  &  0.218 & 0.00e+0  \\
  no emission          &   -19.678      &   -20.113   & 0.435  &  0.343 & 0.00e+0 \\\hline
     total                 &   -19.378     &   -19.888   & 0.510  &  0.306 & 0.00e+0 \\\hline
\multicolumn{6}{c}{\bf volume limited sample} \\\hline
   type I              &   -20.530   &   -20.682  & 0.152  &  0.152 & 1.05e-3  \\
  Seyfert              &   -20.482   &   -20.525  & 0.043  &  0.152 & 6.77e-2  \\       
   LINER               &   -20.497   &   -20.637  & 0.140  &  0.161 & 5.97e-1  \\ \hline
    AGN                &   -20.520   &   -20.630  & 0.110  &  0.138 & 1.13e-4  \\
  composite            &   -20.430   &   -20.501  & 0.071  &  0.093 & 4.01e-7  \\
 star forming          &   -20.362   &   -20.394  & 0.032  &  0.049 & 7.84e-4  \\\hline
emission line galaxies &   -20.405   &   -20.464  & 0.059  &  0.074 & 5.68e-15  \\
  no emission          &   -20.504   &   -20.604  & 0.100  &  0.102 & 1.74e-38  \\\hline
     total             &   -20.459   &   -20.556  & 0.097  &  0.101 & 0.00e+0 \\\hline
\multicolumn{6}{c}{\bf faint sample} \\\hline
   type I              &   -19.615   &   -19.673  & 0.058  &  0.123 & 1.30e-4  \\
  Seyfert              &   -19.552   &   -19.590  & 0.038  &  0.064 & 1.72e-1  \\
   LINER               &   -19.530   &   -19.636  & 0.106  &  0.163 & 8.41e-2  \\\hline
    AGN                &   -19.580   &   -19.632  & 0.052  &  0.081 & 3.17e-4  \\
  composite            &   -19.540   &   -19.594  & 0.054  &  0.048 & 8.77e-8  \\
 star forming          &   -19.037   &   -19.177  & 0.140  &  0.067 & 0.00e+0  \\\hline
emission line galaxies &   -19.069   &   -19.241  & 0.172  &  0.079 & 0.00e+0  \\
  no emission          &   -19.558   &   -19.555  &-0.003  &  0.024 & 1.23e-8  \\\hline
     total             &   -19.261   &   -19.402  & 0.141  &  0.079 & 0.00e+0  \\\hline\hline\\\hline


\hline\hline
\bf sSFR & in voids  & in walls &v-w\tablenotemark{a} &  D\tablenotemark{b}  &  PROB\tablenotemark{c}  \\\hline
\multicolumn{6}{c}{\bf volume limited sample} \\\hline
   type I                    &   -10.259  &  -10.486  &  0.227  &  0.188  & 5.70e-4  \\
  Seyfert                    &   -10.831  &  -10.886  &  0.055  &  0.111  & 3.29e-1  \\
   LINER                     &   -10.642  &  -11.075  &  0.433  &  0.373  & 3.52e-3  \\\hline
    AGN                      &   -10.530  &  -10.718  &  0.188  &  0.141  & 5.19e-4  \\
  composite                  &   -10.321  &  -10.356  &  0.035  &  0.041  & 9.50e-2  \\
 star forming                &    -9.933  &   -9.969  &  0.036  &  0.067  & 6.48e-7  \\\hline
emission line galaxies       &   -10.085  &  -10.162  &  0.077  &  0.081  & 1.10e-17  \\
  no emission                &   -11.547  &  -11.730  &  0.183  &  0.127  & 0.00e+0  \\\hline
     total                   &   -10.695  &  -11.329  &  0.634  &  0.153  & 0.00e+0   \\\hline
\multicolumn{6}{c}{\bf faint sample} \\\hline
   type I                     &   -10.238  &  -10.421   &  0.183  & 0.136  & 2.35e-4  \\
  Seyfert                     &   -10.834  &  -10.868   &  0.034  & 0.052  & 3.96e-1  \\
   LINER                      &   -10.849  &  -10.778   & -0.071  & 0.096  & 6.50e-1  \\\hline
    AGN                       &   -10.686  &  -10.741   &  0.055  & 0.072  & 3.88e-3  \\
  composite                   &   -10.376  &  -10.404   &  0.028  & 0.040  & 1.37e-5  \\
 star forming                 &    -9.836  &   -9.872   &  0.036  & 0.053  & 0.00e+0 \\\hline
emission line galaxies        &    -9.884  &   -9.937   &  0.053  & 0.061  & 0.00e+0  \\
  no emission                 &   -11.256  &  -11.367   &  0.111  & 0.073  & 0.00e+0 \\\hline
     total                    &   -10.077  &  -10.333   &  0.136  & 0.136  & 0.00e+0  \\\hline\hline


\\

     \end{tabular*}

    \tablenotetext{0} {Median values of $M_r$, $\rm log M_*/M_\sun$, D4000, sSFR, and $\rm lg L[O\ {III}]/L_\sun$ and the Kolmogorov-Smirnov test results between void galaxies and wall galaxies.}
    \tablenotetext{1} {Medians of the void galaxy sample minus medians of the wall galaxy sample}
    \tablenotetext{2} {Floating scalar giving the Kolmogorov-Smirnov statistic. It specifies the maximum deviation between the cumulative distribution of the data and the supplied function.}
    \tablenotetext{3} {Floating scalar between 0 and 1, giving the significance level of the K-S statistic. Small values of PROBs corespond to more significant Ds (distances).}

   \end{minipage}

\end{tabular}

\end{table}


\acknowledgments
\noindent {\bf Ackowledgements:}
\vspace{0.2in}

The research presented here is partially supported by the National Natural Science Foundation of China under grants No. 11473305, by the Strategic Priority Research Program  ``The Emergence of Cosmological Structures'' of Chinese Academy of Sciences, Grant No. XDB09030200.

D.P. acknowledge the supports by the Chinese Academy of Sciences, Grant No.Y38528001, and by the National Natural Science Foundation of China International Cooperation and Exchange Program under grants No.Y34556001.



\begin{thebibliography}{}
\expandafter\ifx\csname natexlab\endcsname\relax\def\natexlab#1{#1}\fi

\end{thebibliography}


\begin{thebibliography}{}
\expandafter\ifx\csname natexlab\endcsname\relax\def\natexlab#1{#1}\fi

\bibitem[{{Abazajian} {et~al.}(2009){Abazajian}, {Adelman-McCarthy},
  {Ag{\"u}eros}, {Allam}, {Allende Prieto}, {An}, {Anderson}, {Anderson},
  {Annis}, {Bahcall}, \& et~al.}]{aba09}
{Abazajian}, K.~N., {Adelman-McCarthy}, J.~K., {Ag{\"u}eros}, M.~A., {et~al.}
  2009, \apjs, 182, 543

\bibitem[{{Alexander} \& {Hickox}(2012)}]{Alx12}
{Alexander}, D.~M., \& {Hickox}, R.~C. 2012, \nar, 56, 93

\bibitem[{{Alpaslan} {et~al.}(2014){Alpaslan}, {Robotham}, {Obreschkow},
  {Penny}, {Driver}, {Norberg}, {Brough}, {Brown}, {Cluver}, {Holwerda},
  {Hopkins}, {van Kampen}, {Kelvin}, {Lara-Lopez}, {Liske}, {Loveday},
  {Mahajan}, \& {Pimbblet}}]{Alpaslan2014}
{Alpaslan}, M., {Robotham}, A.~S.~G., {Obreschkow}, D., {et~al.} 2014, \mnras,
  440, L106

\bibitem[{{Antonucci}(1993)}]{antonucci93}
{Antonucci}, R. 1993, \araa, 31, 473

\bibitem[{{Arsenault}(1989)}]{ars89}
{Arsenault}, R. 1989, \aap, 217, 66

\bibitem[{{Athanassoula}(2003)}]{atha03}
{Athanassoula}, E. 2003, \mnras, 341, 1179

\bibitem[{{Athanassoula} {et~al.}(2005){Athanassoula}, {Lambert}, \&
  {Dehnen}}]{atha05}
{Athanassoula}, E., {Lambert}, J.~C., \& {Dehnen}, W. 2005, \mnras, 363, 496

\bibitem[{{Baldwin} {et~al.}(1981){Baldwin}, {Phillips}, \&
  {Terlevich}}]{bpt81}
{Baldwin}, J.~A., {Phillips}, M.~M., \& {Terlevich}, R. 1981, \pasp, 93, 5

\bibitem[{{Balogh} {et~al.}(1999){Balogh}, {Morris}, {Yee}, {Carlberg}, \&
  {Ellingson}}]{bal99}
{Balogh}, M.~L., {Morris}, S.~L., {Yee}, H.~K.~C., {Carlberg}, R.~G., \&
  {Ellingson}, E. 1999, \apj, 527, 54

\bibitem[{{Barnes} \& {Hernquist}(1996)}]{barnes96}
{Barnes}, J.~E., \& {Hernquist}, L. 1996, \apj, 471, 115

\bibitem[{{Barnes} \& {Hernquist}(1991)}]{barnes91}
{Barnes}, J.~E., \& {Hernquist}, L.~E. 1991, \apjl, 370, L65

\bibitem[{{Bell} {et~al.}(2004){Bell}, {Wolf}, {Meisenheimer}, {Rix}, {Borch},
  {Dye}, {Kleinheinrich}, {Wisotzki}, \& {McIntosh}}]{bell04}
{Bell}, E.~F., {Wolf}, C., {Meisenheimer}, K., {et~al.} 2004, \apj, 608, 752

\bibitem[{{Benson} {et~al.}(2003){Benson}, {Hoyle}, {Torres}, \&
  {Vogeley}}]{ben03}
{Benson}, A.~J., {Hoyle}, F., {Torres}, F., \& {Vogeley}, M.~S. 2003, \mnras,
  340, 160

\bibitem[{{Blanton} {et~al.}(2005){Blanton}, {Schlegel}, {Strauss},
  {Brinkmann}, {Finkbeiner}, {Fukugita}, {Gunn}, {Hogg}, {Ivezi{\'c}}, {Knapp},
  {Lupton}, {Munn}, {Schneider}, {Tegmark}, \& {Zehavi}}]{blanton05}
{Blanton}, M.~R., {Schlegel}, D.~J., {Strauss}, M.~A., {et~al.} 2005, \aj, 129,
  2562

\bibitem[{{Bournaud} {et~al.}(2005){Bournaud}, {Combes}, \& {Semelin}}]{bour05}
{Bournaud}, F., {Combes}, F., \& {Semelin}, B. 2005, \mnras, 364, L18

\bibitem[{{Brinchmann} {et~al.}(2004){Brinchmann}, {Charlot}, {White},
  {Tremonti}, {Kauffmann}, {Heckman}, \& {Brinkmann}}]{brinch04}
{Brinchmann}, J., {Charlot}, S., {White}, S.~D.~M., {et~al.} 2004, \mnras, 351,
  1151

\bibitem[{{Bruzual} \& {Charlot}(2003)}]{bc03}
{Bruzual}, G., \& {Charlot}, S. 2003, \mnras, 344, 1000

\bibitem[{{Capetti} \& {Baldi}(2011)}]{cape11}
{Capetti}, A., \& {Baldi}, R.~D. 2011, \aap, 529, A126

\bibitem[{{Cardelli} {et~al.}(1989){Cardelli}, {Clayton}, \&
  {Mathis}}]{cardelli89}
{Cardelli}, J.~A., {Clayton}, G.~C., \& {Mathis}, J.~S. 1989, \apj, 345, 245

\bibitem[{{Charlot} \& {Longhetti}(2001)}]{cl01}
{Charlot}, S., \& {Longhetti}, M. 2001, \mnras, 323, 887

\bibitem[{{Choi} {et~al.}(2010){Choi}, {Han}, \& {Kim}}]{chk10}
{Choi}, Y.-Y., {Han}, D.-H., \& {Kim}, S.~S. 2010, Journal of Korean
  Astronomical Society, 43, 191

\bibitem[{{Cid Fernandes} {et~al.}(2011){Cid Fernandes}, {Stasi{\'n}ska},
  {Mateus}, \& {Vale Asari}}]{cid11}
{Cid Fernandes}, R., {Stasi{\'n}ska}, G., {Mateus}, A., \& {Vale Asari}, N.
  2011, \mnras, 413, 1687

\bibitem[{{Colberg} {et~al.}(2008){Colberg}, {Pearce}, {Foster}, {Platen},
  {Brunino}, {Neyrinck}, {Basilakos}, {Fairall}, {Feldman}, {Gottl{\"o}ber},
  {Hahn}, {Hoyle}, {M{\"u}ller}, {Nelson}, {Plionis}, {Porciani}, {Shandarin},
  {Vogeley}, \& {van de Weygaert}}]{colberg08}
{Colberg}, J.~M., {Pearce}, F., {Foster}, C., {et~al.} 2008, \mnras, 387, 933

\bibitem[{{Constantin} {et~al.}(2008){Constantin}, {Hoyle}, \&
  {Vogeley}}]{co08}
{Constantin}, A., {Hoyle}, F., \& {Vogeley}, M.~S. 2008, \apj, 673, 715

\bibitem[{{Debuhr} {et~al.}(2011){Debuhr}, {Quataert}, \& {Ma}}]{debuhr11}
{Debuhr}, J., {Quataert}, E., \& {Ma}, C.-P. 2011, \mnras, 412, 1341

\bibitem[{{Di Matteo} {et~al.}(2005){Di Matteo}, {Springel}, \&
  {Hernquist}}]{dimatteo05}
{Di Matteo}, T., {Springel}, V., \& {Hernquist}, L. 2005, \nat, 433, 604

\bibitem[{{Eastman} {et~al.}(2007){Eastman}, {Martini}, {Sivakoff}, {Kelson},
  {Mulchaey}, \& {Tran}}]{east07}
{Eastman}, J., {Martini}, P., {Sivakoff}, G., {et~al.} 2007, \apjl, 664, L9

\bibitem[{{Eracleous} {et~al.}(2010){Eracleous}, {Hwang}, \& {Flohic}}]{erac10}
{Eracleous}, M., {Hwang}, J.~A., \& {Flohic}, H.~M.~L.~G. 2010, \apjs, 187, 135

\bibitem[{{Ferrarese} \& {Merritt}(2000)}]{Ferrarese2000}
{Ferrarese}, L., \& {Merritt}, D. 2000, \apjl, 539, L9

\bibitem[{{Gebhardt} {et~al.}(2000){Gebhardt}, {Bender}, {Bower}, {Dressler},
  {Faber}, {Filippenko}, {Green}, {Grillmair}, {Ho}, {Kormendy}, {Lauer},
  {Magorrian}, {Pinkney}, {Richstone}, \& {Tremaine}}]{Gebhardt2000}
{Gebhardt}, K., {Bender}, R., {Bower}, G., {et~al.} 2000, \apjl, 539, L13

\bibitem[{{Goldberg} {et~al.}(2005){Goldberg}, {Jones}, {Hoyle}, {Rojas},
  {Vogeley}, \& {Blanton}}]{MF}
{Goldberg}, D.~M., {Jones}, T.~D., {Hoyle}, F., {et~al.} 2005, \apj, 621, 643

\bibitem[{{Gottl{\"o}ber} {et~al.}(2003){Gottl{\"o}ber}, {{\L}okas}, {Klypin},
  \& {Hoffman}}]{gottl2003}
{Gottl{\"o}ber}, S., {{\L}okas}, E.~L., {Klypin}, A., \& {Hoffman}, Y. 2003,
  \mnras, 344, 715

\bibitem[{{Gregory} \& {Thompson}(1978)}]{kitt78}
{Gregory}, S.~A., \& {Thompson}, L.~A. 1978, \apj, 222, 784

\bibitem[{{Hao} {et~al.}(2009){Hao}, {Jogee}, {Barazza}, {Marinova}, \&
  {Shen}}]{hao09}
{Hao}, L., {Jogee}, S., {Barazza}, F.-D., {Marinova}, I., \& {Shen}, J. 2009,
  419, 402

\bibitem[{{Hao} {et~al.}(2005){Hao}, {Strauss}, {Tremonti}, {Schlegel},
  {Heckman}, {Kauffmann}, {Blanton}, {Fan}, {Gunn}, {Hall}, {Ivezi{\'c}},
  {Knapp}, {Krolik}, {Lupton}, {Richards}, {Schneider}, {Strateva}, {Zakamska},
  {Brinkmann}, {Brunner}, \& {Szokoly}}]{hao05}
{Hao}, L., {Strauss}, M.~A., {Tremonti}, C.~A., {et~al.} 2005, \aj, 129, 1783

\bibitem[{{Heckman} {et~al.}(2004){Heckman}, {Kauffmann}, {Brinchmann},
  {Charlot}, {Tremonti}, \& {White}}]{heck04}
{Heckman}, T.~M., {Kauffmann}, G., {Brinchmann}, J., {et~al.} 2004, \apj, 613,
  109

\bibitem[{{Hernquist}(1989)}]{hernquist89}
{Hernquist}, L. 1989, \nat, 340, 687

\bibitem[{{Ho} {et~al.}(1997{\natexlab{a}}){Ho}, {Filippenko}, \&
  {Sargent}}]{ho1997}
{Ho}, L.~C., {Filippenko}, A.~V., \& {Sargent}, W.~L.~W. 1997{\natexlab{a}},
  \apj, 487, 568

\bibitem[{{Ho} {et~al.}(1997{\natexlab{b}}){Ho}, {Filippenko}, \&
  {Sargent}}]{ho97}
---. 1997{\natexlab{b}}, \apj, 487, 591

\bibitem[{{Hopkins} {et~al.}(2006){Hopkins}, {Hernquist}, {Cox}, {Di Matteo},
  {Robertson}, \& {Springel}}]{hopkins06}
{Hopkins}, P.~F., {Hernquist}, L., {Cox}, T.~J., {et~al.} 2006, \apjs, 163, 1

\bibitem[{{Hopkins} \& {Quataert}(2010)}]{hopkins10}
{Hopkins}, P.~F., \& {Quataert}, E. 2010, \mnras, 407, 1529

\bibitem[{{Hoyle} {et~al.}(2005){Hoyle}, {Rojas}, {Vogeley}, \&
  {Brinkmann}}]{LF}
{Hoyle}, F., {Rojas}, R.~R., {Vogeley}, M.~S., \& {Brinkmann}, J. 2005, \apj,
  620, 618

\bibitem[{{Hoyle} \& {Vogeley}(2002)}]{hoyle02}
{Hoyle}, F., \& {Vogeley}, M.~S. 2002, \apj, 566, 641

\bibitem[{{Hoyle} {et~al.}(2012){Hoyle}, {Vogeley}, \& {Pan}}]{hoyle12}
{Hoyle}, F., {Vogeley}, M.~S., \& {Pan}, D. 2012, \mnras, 426, 3041

\bibitem[{{Jogee}(2006)}]{jogee06}
{Jogee}, S. 2006, in Lecture Notes in Physics, Berlin Springer Verlag, Vol.
  693, Physics of Active Galactic Nuclei at all Scales, ed. D.~{Alloin}, 143

\bibitem[{{Kauffmann} {et~al.}(2004){Kauffmann}, {White}, {Heckman},
  {M{\'e}nard}, {Brinchmann}, {Charlot}, {Tremonti}, \& {Brinkmann}}]{ka04}
{Kauffmann}, G., {White}, S.~D.~M., {Heckman}, T.~M., {et~al.} 2004, \mnras,
  353, 713

\bibitem[{{Kauffmann} {et~al.}(2003{\natexlab{a}}){Kauffmann}, {Heckman},
  {White}, {Charlot}, {Tremonti}, {Brinchmann}, {Bruzual}, {Peng}, {Seibert},
  {Bernardi}, {Blanton}, {Brinkmann}, {Castander}, {Cs{\'a}bai}, {Fukugita},
  {Ivezic}, {Munn}, {Nichol}, {Padmanabhan}, {Thakar}, {Weinberg}, \&
  {York}}]{ka03stm}
{Kauffmann}, G., {Heckman}, T.~M., {White}, S.~D.~M., {et~al.}
  2003{\natexlab{a}}, \mnras, 341, 33

\bibitem[{{Kauffmann} {et~al.}(2003{\natexlab{b}}){Kauffmann}, {Heckman},
  {Tremonti}, {Brinchmann}, {Charlot}, {White}, {Ridgway}, {Brinkmann},
  {Fukugita}, {Hall}, {Ivezi{\'c}}, {Richards}, \& {Schneider}}]{ka03b}
{Kauffmann}, G., {Heckman}, T.~M., {Tremonti}, C., {et~al.} 2003{\natexlab{b}},
  \mnras, 346, 1055

\bibitem[{{Kewley} {et~al.}(2001){Kewley}, {Dopita}, {Sutherland}, {Heisler},
  \& {Trevena}}]{ke01}
{Kewley}, L.~J., {Dopita}, M.~A., {Sutherland}, R.~S., {Heisler}, C.~A., \&
  {Trevena}, J. 2001, \apj, 556, 121

\bibitem[{{Kewley} {et~al.}(2006){Kewley}, {Groves}, {Kauffmann}, \&
  {Heckman}}]{ke06}
{Kewley}, L.~J., {Groves}, B., {Kauffmann}, G., \& {Heckman}, T. 2006, \mnras,
  372, 961

\bibitem[{{Knapen} {et~al.}(2000){Knapen}, {Shlosman}, \& {Peletier}}]{knap00}
{Knapen}, J.~H., {Shlosman}, I., \& {Peletier}, R.~F. 2000, \apj, 529, 93

\bibitem[{{Kormendy} \& {Ho}(2013)}]{Kormendy_Ho2013}
{Kormendy}, J., \& {Ho}, L.~C. 2013, \araa, 51, 511

\bibitem[{{Laine} {et~al.}(2002){Laine}, {Shlosman}, {Knapen}, \&
  {Peletier}}]{laine02}
{Laine}, S., {Shlosman}, I., {Knapen}, J.~H., \& {Peletier}, R.~F. 2002, \apj,
  567, 97

\bibitem[{{Laurikainen} {et~al.}(2004){Laurikainen}, {Salo}, \&
  {Buta}}]{laur04}
{Laurikainen}, E., {Salo}, H., \& {Buta}, R. 2004, \apj, 607, 103

\bibitem[{{Lee} {et~al.}(2012){Lee}, {Park}, {Lee}, \& {Choi}}]{lee12}
{Lee}, G.-H., {Park}, C., {Lee}, M.~G., \& {Choi}, Y.-Y. 2012, \apj, 745, 125

\bibitem[{{Lehmer} {et~al.}(2009){Lehmer}, {Alexander}, {Chapman}, {Smail},
  {Bauer}, {Brandt}, {Geach}, {Matsuda}, {Mullaney}, \& {Swinbank}}]{lehmer09}
{Lehmer}, B.~D., {Alexander}, D.~M., {Chapman}, S.~C., {et~al.} 2009, \mnras,
  400, 299

\bibitem[{{Martini} {et~al.}(2006){Martini}, {Kelson}, {Kim}, {Mulchaey}, \&
  {Athey}}]{martini06}
{Martini}, P., {Kelson}, D.~D., {Kim}, E., {Mulchaey}, J.~S., \& {Athey}, A.~A.
  2006, \apj, 644, 116

\bibitem[{{Martini} {et~al.}(2009){Martini}, {Sivakoff}, \&
  {Mulchaey}}]{martini09}
{Martini}, P., {Sivakoff}, G.~R., \& {Mulchaey}, J.~S. 2009, \apj, 701, 66

\bibitem[{{McLeod} \& {Rieke}(1995)}]{mcl95}
{McLeod}, K.~K., \& {Rieke}, G.~H. 1995, \apj, 441, 96

\bibitem[{{Moles} {et~al.}(1995){Moles}, {Marquez}, \& {Perez}}]{moles95}
{Moles}, M., {Marquez}, I., \& {Perez}, E. 1995, \apj, 438, 604

\bibitem[{{Mulchaey} \& {Regan}(1997)}]{mu97}
{Mulchaey}, J.~S., \& {Regan}, M.~W. 1997, \apjl, 482, L135

\bibitem[{{Osterbrock} \& {Ferland}(2006)}]{oster06}
{Osterbrock}, D.~E., \& {Ferland}, G.~J. 2006, {Astrophysics of gaseous nebulae
  and active galactic nuclei}

\bibitem[{{Pan} {et~al.}(2012){Pan}, {Vogeley}, {Hoyle}, {Choi}, \&
  {Park}}]{pan12}
{Pan}, D.~C., {Vogeley}, M.~S., {Hoyle}, F., {Choi}, Y.-Y., \& {Park}, C. 2012,
  \mnras, 421, 926

\bibitem[{{Park} {et~al.}(2007){Park}, {Choi}, {Vogeley}, {Gott}, {Blanton}, \&
  {SDSS Collaboration}}]{park07}
{Park}, C., {Choi}, Y.-Y., {Vogeley}, M.~S., {et~al.} 2007, \apj, 658, 898

\bibitem[{{Ricciardelli} {et~al.}(2014){Ricciardelli}, {Cava}, {Varela}, \&
  {Quilis}}]{ricc14}
{Ricciardelli}, E., {Cava}, A., {Varela}, J., \& {Quilis}, V. 2014, \mnras,
  445, 4045

\bibitem[{{Richards} {et~al.}(2002){Richards}, {Fan}, {Newberg}, {Strauss},
  {Vanden Berk}, {Schneider}, {Yanny}, {Boucher}, {Burles}, {Frieman}, {Gunn},
  {Hall}, {Ivezi{\'c}}, {Kent}, {Loveday}, {Lupton}, {Rockosi}, {Schlegel},
  {Stoughton}, {SubbaRao}, \& {York}}]{rich2002}
{Richards}, G.~T., {Fan}, X., {Newberg}, H.~J., {et~al.} 2002, \aj, 123, 2945

\bibitem[{{Rojas} {et~al.}(2004){Rojas}, {Vogeley}, {Hoyle}, \&
  {Brinkmann}}]{rojas04}
{Rojas}, R.~R., {Vogeley}, M.~S., {Hoyle}, F., \& {Brinkmann}, J. 2004, \apj,
  617, 50

\bibitem[{{Rojas} {et~al.}(2005){Rojas}, {Vogeley}, {Hoyle}, \&
  {Brinkmann}}]{rojas05}
---. 2005, \apj, 624, 571

\bibitem[{{Salim} {et~al.}(2007){Salim}, {Rich}, {Charlot}, {Brinchmann},
  {Johnson}, {Schiminovich}, {Seibert}, {Mallery}, {Heckman}, {Forster},
  {Friedman}, {Martin}, {Morrissey}, {Neff}, {Small}, {Wyder}, {Bianchi},
  {Donas}, {Lee}, {Madore}, {Milliard}, {Szalay}, {Welsh}, \& {Yi}}]{salim07}
{Salim}, S., {Rich}, R.~M., {Charlot}, S., {et~al.} 2007, \apjs, 173, 267

\bibitem[{{Sanders} {et~al.}(1988){Sanders}, {Soifer}, {Elias}, {Madore},
  {Matthews}, {Neugebauer}, \& {Scoville}}]{sanders88}
{Sanders}, D.~B., {Soifer}, B.~T., {Elias}, J.~H., {et~al.} 1988, \apj, 325, 74

\bibitem[{{Sarzi} {et~al.}(2010){Sarzi}, {Shields}, {Schawinski}, {Jeong},
  {Shapiro}, {Bacon}, {Bureau}, {Cappellari}, {Davies}, {de Zeeuw}, {Emsellem},
  {Falc{\'o}n-Barroso}, {Krajnovi{\'c}}, {Kuntschner}, {McDermid}, {Peletier},
  {van den Bosch}, {van de Ven}, \& {Yi}}]{sarzi10}
{Sarzi}, M., {Shields}, J.~C., {Schawinski}, K., {et~al.} 2010, \mnras, 402,
  2187

\bibitem[{{Schlegel} {et~al.}(1998){Schlegel}, {Finkbeiner}, \&
  {Davis}}]{schlegel98}
{Schlegel}, D.~J., {Finkbeiner}, D.~P., \& {Davis}, M. 1998, \apj, 500, 525

\bibitem[{{Shen} \& {Sellwood}(2004)}]{shen04}
{Shen}, J., \& {Sellwood}, J.~A. 2004, \apj, 604, 614

\bibitem[{{Stasi{\'n}ska} {et~al.}(2008){Stasi{\'n}ska}, {Vale Asari}, {Cid
  Fernandes}, {Gomes}, {Schlickmann}, {Mateus}, {Schoenell}, {Sodr{\'e}}, \&
  {Seagal Collaboration}}]{stas08}
{Stasi{\'n}ska}, G., {Vale Asari}, N., {Cid Fernandes}, R., {et~al.} 2008,
  \mnras, 391, L29

\bibitem[{{Strauss} {et~al.}(2002){Strauss}, {Weinberg}, {Lupton}, {Narayanan},
  {Annis}, {Bernardi}, {Blanton}, {Burles}, {Connolly}, {Dalcanton}, {Doi},
  {Eisenstein}, {Frieman}, {Fukugita}, {Gunn}, {Ivezi{\'c}}, {Kent}, {Kim},
  {Knapp}, {Kron}, {Munn}, {Newberg}, {Nichol}, {Okamura}, {Quinn}, {Richmond},
  {Schlegel}, {Shimasaku}, {SubbaRao}, {Szalay}, {Vanden Berk}, {Vogeley},
  {Yanny}, {Yasuda}, {York}, \& {Zehavi}}]{strauss2002}
{Strauss}, M.~A., {Weinberg}, D.~H., {Lupton}, R.~H., {et~al.} 2002, \aj, 124,
  1810

\bibitem[{{Tremaine} {et~al.}(2002){Tremaine}, {Gebhardt}, {Bender}, {Bower},
  {Dressler}, {Faber}, {Filippenko}, {Green}, {Grillmair}, {Ho}, {Kormendy},
  {Lauer}, {Magorrian}, {Pinkney}, \& {Richstone}}]{tremaine02}
{Tremaine}, S., {Gebhardt}, K., {Bender}, R., {et~al.} 2002, \apj, 574, 740

\bibitem[{{van de Weygaert} \& {van Kampen}(1993)}]{Weygaert1993}
{van de Weygaert}, R., \& {van Kampen}, E. 1993, \mnras, 263, 481

\bibitem[{{Varela} {et~al.}(2012){Varela}, {Betancort-Rijo}, {Trujillo}, \&
  {Ricciardelli}}]{varela12}
{Varela}, J., {Betancort-Rijo}, J., {Trujillo}, I., \& {Ricciardelli}, E. 2012,
  \apj, 744, 82

\bibitem[{{Veilleux} \& {Osterbrock}(1987)}]{vo87}
{Veilleux}, S., \& {Osterbrock}, D.~E. 1987, \apjs, 63, 295

\bibitem[{{Yan} \& {Blanton}(2012)}]{yan12}
{Yan}, R., \& {Blanton}, M.~R. 2012, \apj, 747, 61

\bibitem[{{York} {et~al.}(2000){York}, {Adelman}, {Anderson}, {Anderson},
  {Annis}, {Bahcall}, {Bakken}, {Barkhouser}, {Bastian}, {Berman}, {Boroski},
  {Bracker}, {Briegel}, {Briggs}, {Brinkmann}, {Brunner}, {Burles}, {Carey},
  {Carr}, {Castander}, {Chen}, {Colestock}, {Connolly}, {Crocker}, {Csabai},
  {Czarapata}, {Davis}, {Doi}, {Dombeck}, {Eisenstein}, {Ellman}, {Elms},
  {Evans}, {Fan}, {Federwitz}, {Fiscelli}, {Friedman}, {Frieman}, {Fukugita},
  {Gillespie}, {Gunn}, {Gurbani}, {de Haas}, {Haldeman}, {Harris}, {Hayes},
  {Heckman}, {Hennessy}, {Hindsley}, {Holm}, {Holmgren}, {Huang}, {Hull},
  {Husby}, {Ichikawa}, {Ichikawa}, {Ivezi{\'c}}, {Kent}, {Kim}, {Kinney},
  {Klaene}, {Kleinman}, {Kleinman}, {Knapp}, {Korienek}, {Kron}, {Kunszt},
  {Lamb}, {Lee}, {Leger}, {Limmongkol}, {Lindenmeyer}, {Long}, {Loomis},
  {Loveday}, {Lucinio}, {Lupton}, {MacKinnon}, {Mannery}, {Mantsch}, {Margon},
  {McGehee}, {McKay}, {Meiksin}, {Merelli}, {Monet}, {Munn}, {Narayanan},
  {Nash}, {Neilsen}, {Neswold}, {Newberg}, {Nichol}, {Nicinski}, {Nonino},
  {Okada}, {Okamura}, {Ostriker}, {Owen}, {Pauls}, {Peoples}, {Peterson},
  {Petravick}, {Pier}, {Pope}, {Pordes}, {Prosapio}, {Rechenmacher}, {Quinn},
  {Richards}, {Richmond}, {Rivetta}, {Rockosi}, {Ruthmansdorfer}, {Sandford},
  {Schlegel}, {Schneider}, {Sekiguchi}, {Sergey}, {Shimasaku}, {Siegmund},
  {Smee}, {Smith}, {Snedden}, {Stone}, {Stoughton}, {Strauss}, {Stubbs},
  {SubbaRao}, {Szalay}, {Szapudi}, {Szokoly}, {Thakar}, {Tremonti}, {Tucker},
  {Uomoto}, {Vanden Berk}, {Vogeley}, {Waddell}, {Wang}, {Watanabe},
  {Weinberg}, {Yanny}, {Yasuda}, \& {SDSS Collaboration}}]{york2000}
{York}, D.~G., {Adelman}, J., {Anderson}, Jr., J.~E., {et~al.} 2000, \aj, 120,
  1579

\end{thebibliography}

\end{document}